%% file: main.tex
\documentclass[11pt,a4paper]{article}
\pdfoutput=1

\input{packages.tex}
\input{macros.tex}

\makeatletter
\def\@fpheader{\relax}
\makeatother

\title{Foreground biases in strong gravitational lensing}

\author[a, b]{Daniel Johnson,}

\affiliation[a]{Laboratoire Univers et Particules de Montpellier (LUPM), 
CNRS \& Université de Montpellier (UMR-5299),
Parvis Alexander Grothendieck, F-34095 Montpellier Cedex 05, France}

\affiliation[b]{Department of Astronomy, University of Cape Town, 7701 Rondebosch,
Cape Town, South Africa}

\emailAdd{daniel.johnson@umontpellier.fr}

\author[a, c]{Pierre Fleury,}

\affiliation[c]{Universit\'{e} Paris-Saclay, CNRS, CEA, Institut de physique th\'{e}orique, 91191, Gif-sur-Yvette, France}

\emailAdd{pierre.fleury@lupm.in2p3.fr}

\author[a]{Julien Larena,}

\emailAdd{julien.larena@umontpellier.fr}

\author[d, e]{Lucia Marchetti}

\affiliation[d]{Inter-University Institute for Data Intensive Astronomy, Department of Astronomy, University of Cape Town, 7701 Rondebosch, Cape Town, South Africa}

\affiliation[e]{INAF - Istituto di Radioastronomia, via Gobetti 101, I-40129 Bologna, Italy}

\emailAdd{lucia.marchetti@uct.ac.za}

\abstract{Strong gravitational lensing is a competitive tool to probe the dark matter and energy content of the Universe. However, significant uncertainties can arise from the choice of lens model, and in particular the parameterisation of the line of sight. In this work, we consider the consequences of ignoring the contribution of foreground perturbers in lens modelling. We derive the explicit form of the degeneracy between the foreground shear and the ellipticity of a power law lens, which renders the former quantity effectively unmeasurable from strong lensing observables, and biases measurements of the latter by a few percent. Nonetheless, we demonstrate that this degeneracy does not affect measurements of the Einstein radius. Foreground tidal effects are also not expected to bias the slope of the potential, and any biases in this slope should not affect the recovery of the Hubble constant. The foreground convergence term adds an additional uncertainty to the measurement of $H_0$, and we show that this uncertainty will be on the order of $1\%$ for lensing systems located along random lines of sight. There is evidence to indicate that the probability of strong lensing is higher towards overdense lines of sight, and this could result in a small systematic bias towards overestimations of $H_0$.}


\date{\today}

\begin{document}

\maketitle
\flushbottom

\section{Introduction}
\label{sec:introduction}

The expansion of the Universe is neatly summarised in the Hubble parameter $H(t)$, which governs the rate of change of the scale factor $a$ and is anchored by $H_0$, the value of $H(t)$ today. Well-constrained measurements of $H_0$ would serve as a crucial piece of the puzzle that is the accelerating cosmological expansion, and are therefore considered amongst the most important goals of this precision era of cosmology. 

Conventional measurements of $H_0$ are broadly divided into ``late Universe'' measurements from the distance ladder, and ``early Universe'' measurements from temperature anisotropies in the cosmic microwave background. However, early and late Universe measurements disagree by $\sim5\sigma$ \cite{Abbot_2018,Cuceu_2019,Planck_2020,Riess_2022}, a discrepancy known as the Hubble tension, which is widely considered to pose one of the largest challenges to the standard model of modern cosmology \cite{Di_Valentino_2021,Freedman_2021,Abdalla_2022}. In light of this ``cosmological crisis'', the need for independent probes has never been stronger, and strong lensing has emerged as a compelling candidate for novel cosmography \cite{Keeton_1997,Oguri_2007,Suyu_2010,Yildirim_2021,Birrer_2022,Treu_2022}. 

The use of strong lensing time delays for the measurement of $H_0$ was first proposed by Refsdal in 1964 \cite{Refsdal_1964}. The product between the Hubble constant and the relative time delays between images ($H_0\Delta t$) depends only on the lens model and the relative positions between which these delays are calculated, and thus a measurement of $\Delta t$ from a single well-modelled system can theoretically provide a well-constrained $H_0$ measurement \cite{Refsdal_1964,Refsdal_1966}.  In recent studies, these time delays have been measured with the tightest reported uncertainties of the order of 1\% \cite{Bonvin_2016,Millon_2020,Kumar_2015,Tewes_2013,Chen_2019,Suyu_2013}. 

Despite its promise, time-delay cosmography is faced with serious challenges. Certain properties of the lens, source and cosmology are largely degenerate with one another, and cannot be precisely determined from lensing observables alone \cite{Gorenstein_1988,Saha_2000,Wucknitz_2002,Liesenborgs_2012}. Of these, perhaps the most important, notorious and well-studied is the mass-sheet transformation (MST) \cite{Falco_1985,Gorenstein_1988,Saha_2000,Schneider_2013}. This transformation arises from the fact that a lens with an additional homogeneous sheet of mass cannot be readily distinguished from another with a modified density slope. While the effect of this transformation on deflection angles cancels out with an unobservable source-position transformation, the $H_0\Delta t$ factor measured in time-delay cosmography is changed, and without independent constraints on the mass profile of the lens, $H_0$ cannot be accurately known \cite{Wucknitz_2002}. The MST is widely considered to be the most serious obstacle to the goal of constraining $H_0$ within 1\% \cite{Keeton_2004,Fassnacht_2011, Kochanek_2020}.

The MST manifests as an uncertainty in both the main lens mass and the mass distributed along the line of sight. To mitigate the latter, several techniques have been proposed and implemented. Constraints on the line-of-sight convergence are possible via estimates of the large scale structure in the vicinity of the lens \cite{Fassnacht_2006, Momcheva_2006,Williams_2006,Auger_2007,Wilson_2017,Suyu_2010,Fassnacht_2011,Springel_2005,Greene_2013,Birrer_2017,Fischer_1997,Nakajima_2009,Fadely_2010,Tihhonova_2018}. While estimates from weak lensing shear measurements in the proximity of the lens can help to add constraints, they can also introduce a bias into the measurement, given the different weighting functions of the convergence terms in weak and strong lensing (see \cite{Fleury_2021a}, for example). This problem also affects estimates from galaxy counts or mass fitting nearby groups, as the contribution of foreground mass will be underestimated relative to that in the background of the lens if the convergence is not properly parameterised. This motivated the work of \cite{McCully_2017}, in which the tidal multi-plane formalism introduced in \cite{McCully_2014} to properly capture these non-linear effects was shown to produce more reliable results from mock data than those which did not account for the effects of foreground perturbers.

The most important and frequently used method for lifting the mass-sheet degeneracy makes use of velocity dispersion data. Some recent examples of this include \cite{Rusu_2019,Birrer_2019,Chen_2019,Wong_2019,Millon_2020,Chen_2011,Shajib_2023}. However, in each of these studies, the convergence was parameterised as a single $\kappa_\mathrm{ext}$ term meant to absorb the effects of both the foreground and the background in the lens system. Reference~\cite{Teodori_2022} points out that this can lead to non-negligible biases in the inferred convergence, as, in general, weak lensing effects manifest differently in kinematics and imaging. This observation is also made in \cite{Gomer_2021}, which argues that lensing degeneracies can appear in the process of constraining the external convergence, biasing the recovery of $H_0$. 

This paper is motivated by the work of \cite{McCully_2017,Fleury_2021a,Etherington_2023}, which demonstrated that perturbative weak lensing effects on strong lensing images cannot be fully absorbed by a single parameter without a sufficiently complex lens model, and in particular by that of \cite{Teodori_2022}, which demonstrated the potential for introducing biases in $H_0$ measurements when under-parameterising these weak lensing terms. We expand on this work by investigating the influence of foreground shear, and derive the first-order degeneracy between this shear and the ellipticity of an elliptical power law lens, demonstrating that neither can be tightly constrained from lensing variables alone. We demonstrate that the measured Einstein radius is unaffected by this degeneracy, and thus $H_0$ inferences should be unbiased by the foreground shear, at least to first order. We also estimate the bias introduced by the foreground convergence and find that it can add an additional uncertainty of up to a percent to $H_0$ measurements for typical lines of sight.

This paper is organised as follows: in \cref{sec:theory}, we summarise the formalism introduced in \cite{Fleury_2021a} to parameterise weak lensing effects on strong lensing images. In \cref{sec:shear} we consider the effects of the foreground shear, deriving the explicit form of the first-order degeneracy between the foreground shear and the ellipticity of an elliptical power law lens, quantifying the strength of this effect, and showing how the Einstein radius of the lens escapes this degeneracy. In \cref{sec:msd_los} we consider the effects of foreground convergence, reviewing the mass-sheet degeneracy, demonstrating the biases introduced into measurements of $H_0$ when constraining the lens mass from velocity dispersion measurements without adequately parameterised weak lensing effects, and presenting the expected strength of this bias. We show, however, that the power law slope is unaffected. We summarise our results in \cref{sec:conclusion}.

\section{Theoretical background}
\label{sec:theory}

\subsection{Line-of-sight perturbations and the minimal model}
At the heart of the study of gravitational lensing is the lens equation, which relates the position in the sky at which a light ray is observed, $\btheta$, to its point of origin in the source plane $\bbeta$. In the presence of a single strong lens plane with lens potential $\psi$ and purely tidal perturbations from matter homogeneities along the line of sight, and ignoring terms equating to unobservable source position transformations, this equation takes the form
\begin{equation}
    \bbeta=\bA_{\os}\btheta-\bA_{\ds}\frac{\rd\psi(\bA_{\od}\btheta)}{\rd\btheta},
    \label{eq:DL_lens_tidal_beta'}
\end{equation}
where the amplification matrices characterise tidal distortions to an infinitesimal beam of light, resulting from foreground perturbers between the observer and lens ($\bA_{\od}$), background perturbers between lens and source ($\bA_{\ds}$) and perturbers between observer and source ($\bA_{\os}$), which is equivalent to the weak lensing amplification matrix in the absence of the main lens. Reference~\cite{Fleury_2021a} can be referred to for a derivation of \cref{eq:DL_lens_tidal_beta'} and for precise expressions of these matrices in terms of the non-dominant lens planes.

\begin{figure}
    \centering
    \includegraphics[]{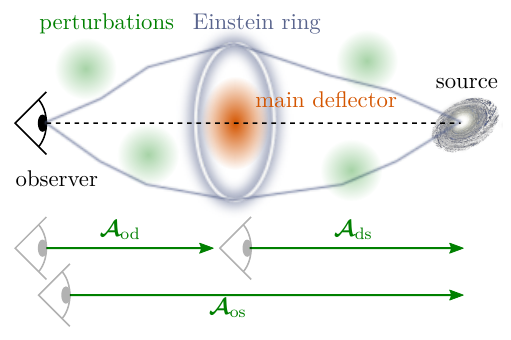}
    \caption{Physical interpretation of $\bA_{\od}$, $\bA_{\ds}$ and $\bA_{\os}$.}
    \label{fig:los_effects}
\end{figure}

Ignoring rotations, these amplification matrices are often parameterised in terms of real convergences $\kappa_{ij}$, corresponding to a surface mass density in units of the critical density and symmetrically stretching an image, and complex shears $\gamma_{ij}=\gamma_1^{ij}+\ri\gamma_2^{ij}$, corresponding to the off-diagonal elements of the matrix which introduce non-symmetric distortions to an image. The amplification matrices can then be written as
\begin{equation}
    \bA_{ij} = \begin{pmatrix} 1-\kappa_{ij}-\gamma_1^{ij}& -\gamma_2^{ij} \\ -\gamma_2^{ij} & 1-\kappa_{ij} + \gamma_1^{ij} \end{pmatrix}.
    \label{eq:a_matrix}
\end{equation}
It is also sometimes convenient to refer to the reduced shear $g_{ij} = (1-\kappa_{ij})^{-1}\gamma_{ij}$ and the reduced amplification matrix $\bG_{ij} = (1-\kappa_{ij})^{-1}\bA_{ij}$.

The source position $\bbeta$ is not directly observable, and thus any transformation of the lens equation \cref{eq:DL_lens_tidal_beta'} which relates the observed image positions $\btheta$ to said transformation of $\bbeta$ is no less applicable to lensing observations. Multiplying the lens equation~\eqref{eq:DL_lens_tidal_beta'} throughout by  $\bA_{\od}\bA_{\ds}^{-1}$ gives
\begin{equation}
    \tilde{\bbeta} = \ALOS\btheta-\frac{\rd \psi_{{\rm eff}}}{\rd \btheta},
    \label{eq:minimal_lens}
\end{equation}
where
\begin{gather}
    \tilde{\bbeta}\equiv \bA_{\od}\bA_{\ds}^{-1}\bbeta,  \\
    \psi_{{\rm eff}}(\btheta)\equiv\psi[\bA_{\od}\btheta],  \\
    \ALOS \equiv \bA_{\od}\bA_{\ds}^{-1}\bA_{\os}.
    \label{eq:source_position_transform_definitions}
\end{gather}
As in \cref{eq:a_matrix}, we can parameterise $\ALOS$ in terms of the line-of-sight convergence $\kappa_\mathrm{LOS}$ and shear $\gamma_\mathrm{LOS}$, from which we can also define the reduced shear $g_\mathrm{LOS}$. 

In \cite{Fleury_2021a}, \cref{eq:minimal_lens} is referred to as the ``minimal lens model'', in that it fully encodes the relevant degeneracies of lensing with tidal perturbations along the line of sight. We see that the amplification matrices $\bA_{\od}$, $\bA_{\os}$ and $\bA_{\ds}$ all contribute to $\ALOS$, and thus they are not measurable independently of one another. This degeneracy is further illustrated in mock data in \cite{Hogg_2023}. While the argument of the lensing potential $\psi$ needs to be evaluated at a position which is modified by the foreground perturbers, a sufficiently general potential $\psi_{\rm eff}$ would be able to accommodate these modifications without the additional term, and thus $\bA_{\od}$ is degenerate with parameters within such a potential. Thus the standard parameterisation of line-of-sight effects in terms of an external convergence and shear is sufficient if and only if the main lens potential is sufficiently flexible.

\subsection{Time delays and the time delay scale}

When line-of-sight effects in the tidal regime are included, the time delay experienced by a light ray relative to an unlensed ray is given by
\begin{equation}
    t(\btheta,\bbeta) = \tau_{\ds}\left[\frac{1}{2}(\btheta-\bbeta')\cdot\bA_{{\rm LOS}}(\btheta-\bbeta')-\psi_{{\rm eff}}(\btheta) \right],
    \label{eq:time_delay_tidal_general}
\end{equation}
where $\bbeta' \equiv \bA_{\os}^{-1}\bbeta$ and $\tau_{\ds}$ is the time delay scale, given by
\begin{equation}
    \tau_{\ds} \equiv (1+z_{\rm d})\frac{D_{\od}D_{\os}}{c D_{\ds}}.
    \label{eq:time_delay_scale}
\end{equation}
While the quantity $t(\btheta,\bbeta)$ is unmeasurable, relative time delays $\Delta t$ between two images (with the same $\bbeta$s but different $\btheta$s) are, and are proportional to $\tau_{\ds}$. In turn, the ratio of angular diameter distances in \cref{eq:time_delay_scale}, when expressed as a function of the lens and source redshifts, ensure that $\tau_{\ds}$ is inversely proportional to $H_0$, and so
\begin{equation}
    \Delta t \propto \frac{1}{H_0}.
\end{equation}
Thus time delay measurements combined with imaging constraints on the effective potential and line-of-sight effects can provide a measurement of $H_0$.

\section{Foreground shear effects}
\label{sec:shear}

As is made clear by the minimal lens model~\eqref{eq:minimal_lens}, the foreground shear matrix appears within the argument of the main lens' potential, and its effects are therefore generally impossible to disentangle from those of the main lens. In particular, the effects of foreground shear are, at least to linear order, degenerate with the effects of the main lens ellipticity. This degeneracy has been demonstrated explicitly for the singular isothermal ellipse (SIE) model in \cite{Fleury_2021a}, but in the following, it will be demonstrated for the more general and more commonly used family of elliptical power law lenses (\cite{Barkana_1998,Tessore_2015}, used in \cite{Wong_2016,Birrer_2016,Bonvin_2016,Birrer_2018,Rusu_2019,Wong_2019,Chen_2019,Shajib_2023} and others). 

In the following, it will prove convenient to represent vector quantities in terms of complex numbers, whereby
\begin{equation}
    \btheta = \theta_1\bm{e}_1+\theta_2\bm{e}_2\mapsto \utheta=\theta_1+\ri \theta_2.
\end{equation}
In the standard parameterisation of an amplification matrix $\bA_{ij}$, the result of multiplying a vector $\bm{u}$ by $\bA_{ij}$ in complex notation is simply 
\begin{equation}
    \bm{v}=\bA_{ij}\bm{u}\mapsto\underline{v}=(1-\kappa_{ij})\underline{u}-\gamma_{ij}\underline{u}^*.
    \label{eq:complex_equivalence}
\end{equation}
    \subsection{Elliptical power laws}

The most commonly used parameterised lens models fall into the class of elliptical power laws, which add a further degree of flexibility to the singular isothermal ellipse \cite{Barkana_1998,Tessore_2015}.\footnote{The SPEMD model presented in \cite{Barkana_1998} and used throughout earlier time-delay cosmography studies is equivalent to this, except that it adds a core radius to the denominator of \cref{eq:EPL_convergence}. The conclusions of this work are applicable to this parameterisation, but we will assume a core radius of 0 and follow the more popular EPL parameterisation of \cite{Tessore_2015}.} The classical parameterisation of this deflector \cite{Tessore_2015} has a convergence which is given by
\begin{equation}
    \kappa(\zeta) = \frac{3-\gamma}{2}\left(\frac{b}{\zeta}\right)^{\gamma-1},
    \label{eq:EPL_convergence}
\end{equation}
where $\gamma$ is the negative power-law slope of the 3D mass distribution, $b$ is a scaling parameter, and $\zeta$ is the elliptical radius, given by
\begin{equation}
    \zeta \equiv \theta\sqrt{q^2\cos^2(\varphi')+\sin^2(\varphi')},
    \label{eq:zeta}
\end{equation}
where $q$ is the minor-to-major axis ratio and $\varphi'=\varphi-\varphi_0$ is the angle measured relative to the axes of the ellipse, with $\varphi=\mathrm{atan}2(\theta_2,\theta_1)$ being the usual polar angle and $\varphi_0$ the inclination of the ellipse.\footnote{Here, we use the two-argument $\mathrm{atan}2(y,x)$ function which returns an angle respecting the quadrant of $x$ and $y$.} This version is generalised from that in \cite{Tessore_2015}, as we do not assume that the ellipticity is aligned with the optical axis. As in the case of the convergences appearing in \cref{eq:a_matrix}, $\kappa(\zeta)$ represents a surface mass density in units of the critical density $\Sigma_\mathrm{cr},$
\begin{equation}
    \Sigma_\mathrm{cr} =\frac{c^2D_{\os}}{4\pi G D_\ds D_\od}.
\end{equation}

Several equivalent reparameterisations of \cref{eq:EPL_convergence,eq:zeta} are possible. As written above, the value of $\zeta$ fixes the major axis across different choices of $q$, while in the parameterisation in \cite{Barkana_1998}, the minor axis is fixed. While the definition in \cite{Barkana_1998} was used in early $H_0$ measurements from strong lensing (such as \cite{Suyu_2010}), later H0LiCOW \cite{Wong_2016,Bonvin_2016,Birrer_2018,Rusu_2019,Wong_2019}, TDCOSMO \cite{Birrer_2020,Shajib_2023} and other $H_0$ measurements \cite{Suyu_2013,Birrer_2016,Chen_2019} redefined $\zeta$ as
\begin{equation}
    \zeta' = \sqrt{q\cos^2(\varphi')+\frac{1}{q}\sin^2(\varphi')},
    \label{eq:zeta_2}
\end{equation}
such that this elliptical radius now corresponds to a fixed average radius across different values of $q$, while simultaneously replacing $b\rightarrow \tthetaE\sqrt{q}$. In the following, we will discuss the physical interpretation of these quantities and will see that this parameterisation most naturally corresponds to the observable properties of the lens.

For the elliptical power law given in \cref{eq:EPL_convergence}, the displacement angle $\balpha = \frac{\rd \psi(\btheta)}{\rd \btheta}$ can be written in complex notation as
\begin{equation}
    \ualpha(\zeta,\varphi)=\frac{2b}{1+q}\left(\frac{b}{\zeta}\right)^{\gamma-2}\mathrm{e}^{\mathrm{i}\varphi'_\mathrm{ell}}\:_2F_1\left(1;\frac{\gamma-1}{2};\frac{3-\gamma}{2};-\frac{1-q}{1+q}\mathrm{e}^{\mathrm{i}2\varphi'_\mathrm{ell}}\right),
    \label{eq:epl_deflection_full}
\end{equation}
where $_2F_1(a;b;c;z)$ is the Gaussian hypergeometric function, and $\varphi_\mathrm{ell}'$ is the elliptical angle, defined by\footnote{Or equivalently as
    \begin{equation}
        \varphi_\mathrm{ell}' = \mathrm{atan}2\left[\theta_2,q\theta_1\right].
    \end{equation}}
\begin{equation}
    \re^{\ri\varphi_\mathrm{ell}} = \frac{q\cos\varphi'+\ri\sin\varphi'}{|q\cos\varphi'+\ri\sin\varphi'|}.
    \label{eq:elliptical_angle}
\end{equation}
The lensing potential (in complex notation) is
\begin{equation}
    \psi(\utheta) = \frac{\theta}{3-\gamma}\frac{\re^{\ri\varphi'}\ualpha^*(\utheta)+\re^{-\ri\varphi'}\ualpha(\utheta)}{2}.
    \label{eq:EPL_potential}
\end{equation}
    \subsection{Approximate forms of the deflection angle and potential}
    \label{sec:azimuthal_degeneracies}

It will be useful to introduce the complex ellipticity $\varepsilon$, the magnitude of which is defined in terms of the axis ratio via  
\begin{equation}
    |\varepsilon| \equiv \frac{1-q^2}{1+q^2}, 
    \label{eq:e_definition}
\end{equation}
such that for small $|\varepsilon|$ we can approximate\footnote{It should be noted that the relationship between $f$ and $\varepsilon$ given in \cite{Fleury_2021a} is slightly different, and would give, to linear order, $f=1-2|\varepsilon|$. The definition of $\varepsilon$ used in \cref{eq:e_definition} is chosen for consistency with the notation in \texttt{lenstronomy} \cite{Birrer_2018}.} $f \approx 1-|\varepsilon|$.  The complex ellipticity is then defined as $\varepsilon \equiv |\varepsilon|\re^{2\ri\varphi_0}$. From here onwards, the ``$\approx$'' symbol will denote a relationship that is true to linear order in $|\varepsilon|$ and the foreground reduced shear, $|g_{\od}|$.

If the axis ratios in \cref{eq:epl_deflection_full} are expressed in terms of $\varepsilon$ and expanded at linear order in $|\varepsilon|$, the deflection angle can be written as
\begin{equation}  
    \ualpha(\utheta) \approx \left(\frac{\theta}{b}\right)^{1-\gamma}\theta\re^{\ri\varphi'}\left\{1+\frac{1}{2}(\gamma-1)|\varepsilon|-\frac{3-\gamma}{2}\mathrm{Re}(\varepsilon\re^{-2\ri\varphi})+\frac{3-\gamma}{5-\gamma}\varepsilon^*\re^{2\ri\varphi}\right\},
    \label{eq:approximate_epl_alpha}
\end{equation}
and so, from \cref{eq:EPL_potential}, the lensing potential can be written as
\begin{equation}
    \psi(\utheta)
    \approx  \left(\frac{\theta}{b}\right)^{1-\gamma}\theta^2\left\{1+\frac{1}{2}(\gamma-1)|\varepsilon|-\frac{1}{2}\frac{3-\gamma}{5-\gamma}\mathrm{Re}(\varepsilon\re^{-2\ri\varphi})\right\}.
    \label{eq:approximate_epl_potential}
\end{equation}
\subsection{Radial dependencies}
\label{sec:radial_dependencies}
Consider the definition of the geometric Einstein radius, which is the angular radius within which the average surface mass density is equal to the critical density.\footnote{The Einstein radius $\thetaE$ is also sometimes defined via the formula $S_\mathrm{crit}=\pi \thetaE^2$, where $S_\mathrm{crit}$ is the area enclosed by the critical curve. For simple axisymmetric lenses, these definitions are equivalent \cite{Cao_2022}.} If the entire lensing mass is well described by an elliptical power law, then we can obtain an expression for the geometric Einstein radius of this lens $\tthetaE$ via
\begin{equation}
    \pi D_\od^2\tthetaE^2\Sigma_\mathrm{cr} = M(<\tthetaE) = D_\od^2\int^{2\pi}_0 \rd\varphi \int^{\tthetaE}_0 \rd\theta \; \theta \kappa(\zeta)\Sigma_\mathrm{cr} . 
    \label{eq:thetaE_integral}
\end{equation}
If we replace $\kappa(\zeta)$ with the expression in \cref{eq:EPL_convergence} and solve for $\tthetaE$, the expression we obtain is
\begin{equation}
    \tilde{\theta}_\mathrm{E} \approx \left(1+\frac{1}{2}|\varepsilon|\right)b.
    \label{eq:epl_geometric_Einstein}
\end{equation}
It should be noted that this is only the geometric Einstein radius of a lens in which the entirety of the lensing convergence is well-described by the power law expression given in \cref{eq:EPL_convergence}. In the presence of an additional external convergence, this would not be the case, and we thus use the notation $\tthetaE$ for the parameter appearing within the EPL model and $\thetaE$ for the observable geometric Einstein radius of the entire lensing mass.

At first glance, \cref{eq:epl_geometric_Einstein} seems to illustrate a degeneracy between the parameters $b$ and $\varepsilon$ when recovering the Einstein radius. From \cref{eq:approximate_epl_potential}, it appears that there is a first-order contribution from $|\varepsilon|$ to the radial structure of the lensing potential, and that, if we were to measure a value of $b$, this contribution would need to be divided out in order to obtain the value of $\tthetaE$. However, the combination in \cref{eq:epl_geometric_Einstein} is precisely the radial component of the lensing potential and deflection angles in \cref{eq:approximate_epl_potential,eq:approximate_epl_alpha} (plus a power of $\gamma-1$). Thus \cref{eq:approximate_epl_potential} can be rewritten as
\begin{equation}
    \psi(\utheta)
    \approx  \left(\frac{\theta}{\tthetaE}\right)^{1-\gamma}\theta^2\left\{1-\frac{1}{2}\frac{3-\gamma}{5-\gamma}\mathrm{Re}(\varepsilon\re^{-2\ri\varphi})\right\},
    \label{eq:epl_potential_with_thetaE}
\end{equation}
and while $\varepsilon$ and $b$ are inextricable, the more relevant parameter $\tthetaE$ is unaffected.

\subsection{The effective potential}

When tidal line-of-sight corrections are included in the lens equation, the potential of the main deflector must be evaluated at a position $\bA_{\od}\btheta$, modified by foreground deflectors, as can be seen in \cref{eq:minimal_lens}. Ignoring the effects of convergence for now, and continuing to work at first order in $g_{\od}$, this modified position can be written as
\begin{equation}
    \utheta \rightarrow  \theta\left(\re^{\ri \varphi} - g_{\od}\re^{-\ri \varphi}\right). 
    \label{eq:modified_position}
\end{equation}
This shear manifests in the magnitude of $\utheta$, via
\begin{equation}
    \theta \rightarrow \theta\left[1-\mathrm{Re}(g_{\od}\re^{-2\ri\varphi}) \right], 
    \label{eq:theta_modified}
\end{equation}
and in the polar angle, giving
\begin{equation}
    \re^{\ri\varphi} \rightarrow \re^{\ri\varphi}\left[1-\ri\,\mathrm{Im}\left(g_{\od}\re^{-2\ri\varphi}\right)\right].
\end{equation}
This angular perturbation contributes only at order $\mathcal{O}(\varepsilon g_{od})$ and higher to deflection angles, and so, simply replacing $\theta$ in \cref{eq:approximate_epl_potential} using \cref{eq:theta_modified}, the effective potential can be expressed as
\begin{equation}
    \psi_{\mathrm{eff}}(\utheta)
    \approx \left(\frac{\theta}{\tthetaE}\right)^{1-\gamma}\theta^2\left(1-(3-\gamma)\mathrm{Re}\left\{\left[\frac{1}{2(5-\gamma)}\varepsilon+g_{\od}\right]\re^{-2\ri\varphi}\right\}\right). 
    \label{eq:azimuthal_degeneracy}
\end{equation}
This result is consistent with that presented in \cite{Fleury_2021a} for the SIE lens, which is recovered by substituting $\gamma=2$ (and noting the different definitions of ellipticity). To linear order, an elliptical power law lens with ellipticity $\varepsilon$ and in the presence of a foreground shear $g_{\od}$ is indistinguishable from a lens without foreground shear and with ellipticity
\begin{equation}
    \varepsilon_\mathrm{eff}\equiv \varepsilon + 2(5-\gamma)g_{\od}. \label{eq:degeneracy_form}
\end{equation}
In the language of \cite{Hogg_2023}, the isopotential contours of the effective potential $\psi_\mathrm{eff}$ are the images of the isopotential contours of the main lens potential $\psi$, distorted by the perturbations encoded in $\bG_{\od}$. Because the ellipticity of the SIE and EPL lens models appear in exact form in the convergence, rather than the potential, this degeneracy is only true to first order for these models. However, a model in which the ellipticity was implemented directly in the lensing potential would render this degeneracy exact, and thus differentiating between $\varepsilon$ and $g_{\od}$ from lensing observables is hopelessly model-dependent. Nonetheless, following the discussion in \cref{sec:radial_dependencies}, the inferred Einstein radius escapes biases arising from this degeneracy, at least to first order.%
    \subsection{Estimating the bias}
    \label{sec:ellipticity_bias}

\begin{figure}
    \centering
    \includegraphics[width=1\columnwidth]{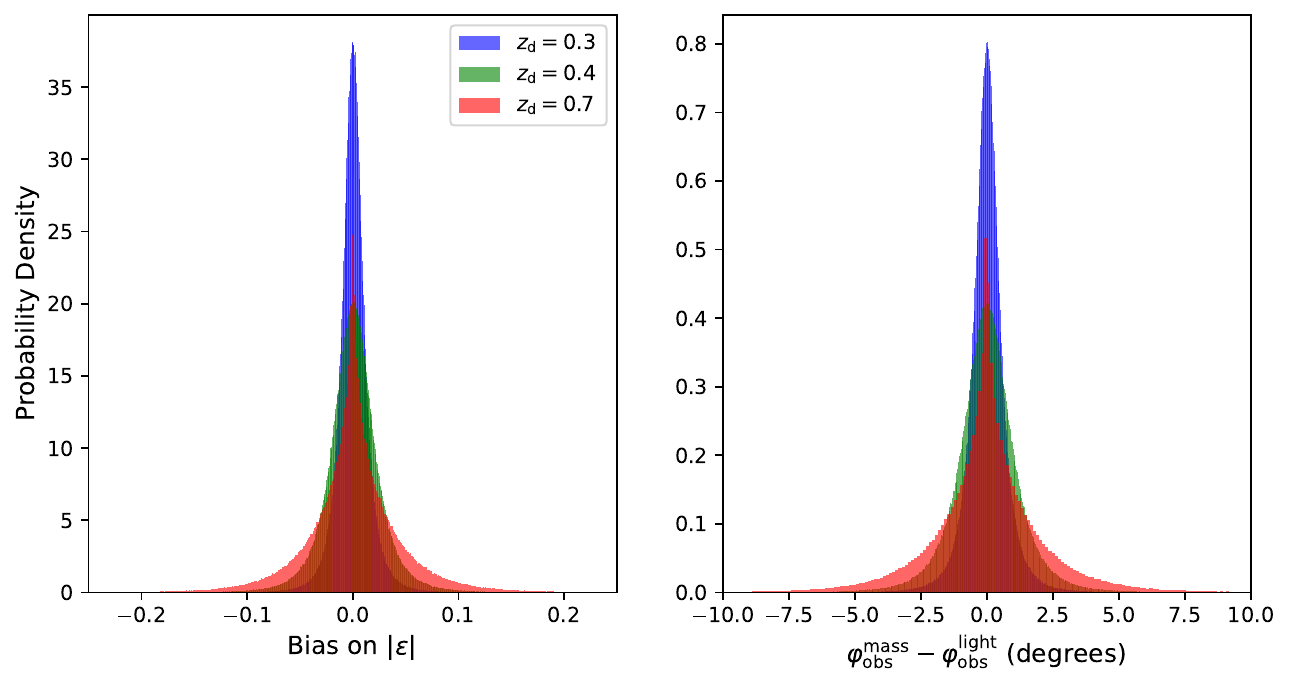}
    \caption{\textit{Left} - the probability density function for $|\varepsilon_\mathrm{eff}|/|\varepsilon|-1$  at different redshifts. The standard deviation for $z=0.3$ is $\sim 2\%$, for $z=0.45$ is $\sim 3\%$ and for $z=0.7$ is $\sim 5\%$. \textit{Right} - the probability density function for $\varphi^\mathrm{mass}_\mathrm{obs}-\varphi^\mathrm{light}_\mathrm{obs}$ for mass and light profiles which are initially aligned, where $\varphi^\mathrm{mass}_\mathrm{obs}$ is the observed inclination of the mass profile, and $\varphi^\mathrm{light}_\mathrm{obs}$ is the observed inclination of the light profile, under the effects of foreground shear. The standard deviation for $z=0.3$ is $\sim 2^\circ$, for $z=0.45$ is $\sim 3^\circ$ and for $z=0.7$ is $\sim 5^\circ$. All distributions have a mean value of 0.}
    \label{fig:ellipticity_histograms}
\end{figure}

For illustration purposes, we present in \cref{fig:ellipticity_histograms} the expected probability distribution of the bias on ellipticity measurements due to the foreground shear, $|\varepsilon_\mathrm{eff}|/|\varepsilon|-1$, for different values of the deflector's redshift~$z_{\rm d}$. We also show the distribution of offsets between the inclination of the mass profile and the lens light profile that would be observed under the effects of the foreground shear, if these profiles were perfectly aligned in the main lens. This offset arises because the effects of foreground shear on the inferred mass profile and the lens light are not identical, with the apparent ellipticity in the light profile simply given by $\varepsilon_\mathrm{obs} = \varepsilon+\gamma_\od$.

To estimate values of $\varepsilon$, we use the results of \cite{Chen_2016}, fitting and sampling from a continuous function to their binned ellipticity distributions. For the power law parameter $\gamma$, we sample from a Gaussian centered on 2, with a standard deviation of 0.05. 

To obtain distributions for $\gamma_\od$ and $g_\od$, we use the results of ray shooting in a dark-matter simulation based on the $N$-body code RAMSES~\cite{teyssier2002cosmological, guillet2011simple}. The simulation uses the WMAP-7 cosmology~\cite{komatsu2011seven},\footnote{The results presented here are expected to remain essentially unchanged if the simulation were performed with a more up-to-date cosmology.} a comoving length of $2625h^{-1}~\text{Mpc}$ and a particle mass of $1.88 \times 10^{10}h^{-1} M_\odot$. Fully relativistic ray tracing has been performed through this simulation by~\cite{Breton:2018wzk} using the \textsc{Magrathea} library~\cite{reverdy2014propagation, Breton}. The resulting HEALPix maps with various weak-lensing quantities, such as convergence, shear and magnification, are publicly available.\footnote{\href{https://cosmo.obspm.fr/public-datasets/raygalgroupsims-relativistic-halo-catalogs}{\tt https://cosmo.obspm.fr/public-datasets/raygalgroupsims-relativistic-halo-catalogs}} 

From this, we find that typical foreground shear values introduce an error on the order of a few percent into the magnitude of the ellipticity, and an artificial difference between the apparent inclinations of the mass and light profiles of a few degrees.

\section{Foreground convergence effects}
\label{sec:msd_los}

Velocity dispersion measurements have played a key role in constraining the lens mass and hence alleviating the mass-sheet degeneracy when measuring $H_0$ with strong lensing. However, the methods with which this has been approached in the last decade can introduce errors into the resulting $H_0$ values. In the following, we will consider errors introduced by neglecting the contribution of foreground convergence terms, as demonstrated in \cite{Teodori_2022}. We will argue that, as with the minimal lens model \cref{eq:minimal_lens}, only $\kappa_{\od}$ and $\kappa_\mathrm{LOS}$ are needed for a minimal parameterisation of the convergence, as $\kappa_{\ds}$ and $\kappa_{\os}$ can be fully absorbed into $\kappa_\mathrm{LOS}$. Nonetheless, omitting $\kappa_{\od}$ leads to biases in the inferred value of $H_0$, for which we will present the estimated probability distribution.

    \subsection{The mass-sheet transformation}

The MST can be simply illustrated as follows. If a new convergence $\kappa_\lambda(\btheta)$ is defined by adding a ``sheet'' of constant mass density $\lambda$ to the lens convergence $\kappa(\btheta)$, while simultaneously rescaling this convergence by\footnote{The roles of $\lambda$ and $1-\lambda$ are sometimes reversed, but both expressions are of course equivalent.} $(1-\lambda)$,
\begin{equation}
    \kappa_\lambda(\btheta) = \lambda+(1-\lambda)\kappa(\btheta),
\end{equation}
the resulting lensing observables are fully degenerate with those of a lens with convergence $\kappa(\btheta)$, differing only by an unobservable source position transformation. This rescaled convergence does not affect relative time delays, but the measured time delay distance and hence the measured Hubble constant is proportional to this MST term as
\begin{equation}
    H_0  \propto \frac{1-\lambda}{\Delta t}.
    \label{eq:H0_msd}
\end{equation}
Thus, without further information to constrain the mass distribution of the main lens, precise measurements of $H_0$ are impossible. 

Now, one important property of tidal line-of-sight corrections is that these effects can be projected and evaluated within the main lens plane. As a consequence, the MST can be reformulated as the unmeasurability of the line-of-sight convergence terms, which, like $\lambda$, correspond to a constant sheet of mass added to the deflector plane. With each of the three line-of-sight convergence terms, we can associate a mass-sheet term which transforms the convergence via $\kappa_{ij} \rightarrow \lambda_{ij}+(1-\lambda_{ij})\kappa_{ij}$. A common choice is to ignore these convergence terms in lens modeling, which implicitly sets
\begin{equation}
    \lambda_{ij} = \frac{\kappa_{ij}}{\kappa_{ij}-1}.
    \label{eq:eliminating_convergences}
\end{equation}
Starting from \cref{eq:time_delay_tidal_general}, it is straightforward to show that $\tau_{\ds}$ is degenerate with
\begin{equation}
    1-\lambda_\mathrm{LOS} \equiv \frac{(1-\lambda_\od)(1-\lambda_\os)}{1-\lambda_\ds},
    \label{eq:lambda_los}
\end{equation}
and so, from \cref{eq:eliminating_convergences,eq:H0_msd}, a value $H_0^\mathrm{model}$ inferred without line-of-sight convergences in the model will be related to the true value of $H_0$ via
\begin{equation}
    H_0 = (1-\kappa_\mathrm{LOS})H_0^{\mathrm{model}}.
    \label{eq:true_correction}
\end{equation}
This is simply a reformulation of the result presented in \cite{Birrer_2020}, and the details  can be found in \cref{appendix}.
    \subsection{Observable Einstein radii and the power law parameter}

In the presence of tidal line-of-sight effects, \cref{eq:minimal_lens} can be rewritten as
\begin{equation}
    \bbeta'=\btheta-\balpha'(\btheta),
\end{equation}
where $\bbeta'$ is the source position (plus an unobservable source position transformation) and 
\begin{equation}
    \balpha'(\btheta) = 2\ALOS^{-1}\frac{\rd \psi(\bA_{\od}\btheta)}{\rd \btheta}.
    \label{eq:modified_alpha}
\end{equation}
As described above, it is typical to make the transformation in \cref{eq:eliminating_convergences} and omit line-of-sight convergences from the model. From \cref{eq:modified_alpha,eq:approximate_epl_alpha}, the observed geometric Einstein radius and the power law scale $\tthetaE$ can be related via \cite{Teodori_2022}
\begin{equation}
    \thetaE =\left[\frac{(1-\kappa_{\od})^{3-\gamma}}{1-\kappa_\mathrm{LOS}}\right]^{\frac{1}{{\gamma-1}}}\tthetaE.
    \label{eq:thetaE_to_thetaE_epl}
\end{equation}
$\tthetaE$ is, once again, a quantity associated with the main lens only, while $\thetaE$ is the geometric Einstein radius of an image which has been distorted
both by the main lens and mass along the line of sight.

    \subsection{Velocity dispersion}
    \label{sec:msd_td}

To constrain the factor relating $\tthetaE$ to $\thetaE$, which has historically been parameterised in terms of $\kappa_\mathrm{ext}$ only, the velocity dispersion of the main lens galaxy must be modelled. Standard practice has been to assume that the mass distribution is a spherically-symmetric power law. The observed Einstein radius is then decomposed in terms of a contribution from this spherical mass distribution and an external convergence, $\kappa_\mathrm{ext}$, which allows the local mass density profile of the main lens to be expressed in terms of $\kappa_\mathrm{ext}$, the observed Einstein radius $\thetaE^\mathrm{fit}$, the power law parameter $\gamma$ and a combination of angular diameter distances. This density profile is then related to the observed velocity dispersion $\sigma(\btheta)$ via spherical Jeans modelling and assumptions on the model of the stellar distribution and anisotropy  \cite{Suyu_2010,Suyu_2013,Wong_2016,Birrer_2018,Rusu_2019,Chen_2019}.

In the assumption of spherical symmetry, there is already a loss of internal consistency, as the lens itself is typically modelled with an elliptical mass profile such as an SIE, an EPL, or a combination of an NFW halo and one or more \Sersic ellipses \cite{Shajib_2023}. Stellar dynamics modelling grows rapidly more complex with departures from spherical symmetry. For triaxial mass distributions, the contribution of the ellipticity to circular velocity is generally an $\mathcal{O}(|\varepsilon|^2)$ correction \cite{Bovy_2023}, and this circular velocity is linearly related to the velocity dispersion \cite{Pizzella_2005,Corsini_2007}. This would suggest that additional errors arising from $|\varepsilon|$ and $g_\od$ in the velocity dispersion measurement to be subdominant compared to the influence of the foreground convergence. The extent to which the projected mass profile and stellar dynamics reflect the true 3D distribution is dependent on inclination angles and stellar anisotropies, but for the purposes of this discussion, we focus on the effects of the external convergence.

Maintaining the spherically symmetric power law assumption and neglecting any stellar anisotropy, the observable velocity dispersion can be written as \cite{Teodori_2022}
\begin{align}
    \sigma^2(\btheta) = 2G\Sigma_{{\rm cr}}D_{\od}\frac{\sqrt{\pi}\Gamma(\frac{\gamma}{2})}{\Gamma(\frac{\gamma-1}{2})}\tthetaE^{\gamma-1}\tilde{\theta}^{2-\gamma}. 
    \label{eq:velocity_dispersion}
\end{align}
In this expression, $\tilde{\theta}=(1-\kappa_{\od})\theta$ is the magnitude of the position at which a ray observed at $\theta$ would have originated from in the main lens plane, omitting any angular effects from foreground shear. As before, $\tthetaE$ refers to a parameter in the EPL main lens and corresponds to the geometric Einstein radius of a lensing system consisting of this lens alone. Adapting the procedure in \cite{Millon_2020,Birrer_2016,Birrer_2019}, we can extract these line-of-sight effects as a prefactor and write
\begin{equation}    
    \sigma^2(\btheta) = \frac{1-\kappa_\mathrm{LOS}}{1-\kappa_{\od}}\frac{D_{\os}}{D_{\ds}}J(\theta,\thetaE,\gamma). 
    \label{eq:velocity_dispersion_2}
\end{equation}
The angular diameter distance terms are introduced by $\Sigma_{\mathrm{crit}}$, and their ratio $D_{\os}/D_{\ds}$ ensures that the Hubble constant factor cancels out. The function $J(\theta,\thetaE^\mathrm{fit},\gamma)$ depends only on lensing observables and is independent of both external convergences and cosmological parameters.

\subsection{The power law slope}

It is worth pausing to note the importance of the power law slope when relating these observables to $\tthetaE$. After the Einstein radius, the leading order observable quantity in lensing is
\begin{equation}
    \frac{D_\theta^2\alpha}{1-D_\theta\alpha}=\frac{D_\theta^3\psi}{1-D_\theta^2\psi},
\end{equation}
where $D_\theta$ is the derivative with respect to the radial variable $\theta$ \cite{Kochanek_2002,Kochanek_2020,Sonnenfeld_2017,Birrer_2021}. From \cref{eq:modified_alpha,eq:thetaE_to_thetaE_epl}, we can express this quantity in terms of the observable $\thetaE$ for a power law lens plus line-of-sight convergence terms as  
\begin{equation}
    \frac{D_\theta^2\alpha}{1-D_\theta\alpha} = \frac{2(3-\gamma)(2-\gamma)(1-\gamma)}{\left(\frac{\theta}{\thetaE}\right)^{\gamma-1}-2(3-\gamma)(2-\gamma)}\left(\frac{1}{\theta}\right).
    \label{eq:radial_observable2}
\end{equation}
Any dependence on the convergence terms is fully absorbed by $\thetaE$, and the resulting expression is simply cubic in $\gamma$ which depends on the observable quantities $\thetaE$ and $D_\theta^2\alpha/(1-D_\theta\alpha)$, and the position $\theta$ at which the latter is evaluated. If we assume that $\gamma$ is close to 2 and evaluate this expression at the Einstein radius, we find 
\begin{equation}
    \gamma \approx 2+\frac{\thetaE}{2}\frac{D_\theta^2\alpha(\thetaE)}{1-D_\theta\alpha(\thetaE)}.
     \label{eq:gamma}
\end{equation}

From \cref{eq:radial_observable2} is it evident that $\gamma$ is not degenerate with $\thetaE$, $\kappa_\od$ or $\kappa_\mathrm{LOS}$ if the main lens is a power law, and so we would expect to measure it without direct systematic biases arising from neglecting these convergence terms in the model. Furthermore, the combination of $\thetaE$ and $\gamma$ which appears in the lens equation (e.g. \cref{eq:epl_potential_with_thetaE}) appears in the exact same form in the expression for the velocity dispersion (\cref{eq:velocity_dispersion}), and thus $H_0$ would remain unaffected as long as the combination $\thetaE^{\gamma-1}$ is well measured. This is, of course, only true if the physical mass distribution is well described by the EPL profile, as biases are known to arise for more physically realistic lenses \cite{Cao_2022}.

\subsection{$H_0$ biases}

In order to constrain the convergence via the velocity dispersion, typically measurements of $\sigma^2(\btheta)$ will be obtained and then compared to predictions of $(D_{\os}/D_{\ds})J(\theta,\thetaE,\gamma)$ from imaging observables (excluding any convergence from the model). The factor by which these terms differed would then be identified as $(1-\kappa_{\mathrm{ext}})$ \footnote{This $\kappa_\mathrm{ext}$ term has historically been used to account not only for matter which is unassociated with the main lens (the ``external'' mass-sheet degeneracy), but also for matter within the main lens which is not well described by the choice of lens profile (the ``internal'' mass-sheet degeneracy). See e.g. \cite{Schneider_2013,Birrer_2022} for a thorough discussion of this.}, and is then used to correct measurements of Hubble constant via
\begin{align}
    H_0^{\mathrm{inferred}} &= (1-\kappa_{\ext}) H_0^{\mathrm{model}}, \\
    &=\frac{1-\kappa_\mathrm{LOS}}{1-\kappa_{\od}}H_0^{\mathrm{model}},
\end{align}
where $H_0^{\mathrm{model}}$ is the value of $H_0$ measured from time delay data with a model without convergence, and $H_0^{\mathrm{inferred}}$ is the value inferred after the correction is applied. However, as seen in \cref{eq:true_correction}, the true correction needed is $(1-\kappa_\mathrm{LOS})$. Therefore, the inferred value of $H_0$ will differ from the true value by a factor of  
\begin{align}
    \frac{H_0^\mathrm{inferred}}{H_0} = \frac{1}{1-\kappa_{\od}}
    \approx 1 + \kappa_{\od}.
\end{align}
Another way to understand this disagreement is in the different ratios angular diameter distances appearing in \cref{eq:time_delay_tidal_general,eq:time_delay_scale} governing time delays and in \cref{eq:velocity_dispersion_2}. The effects of line-of-sight convergence terms can be reformulated as a transformation of angular diameter distances $D_{ij} \rightarrow (1-\kappa_{ij})D_{ij}$, and thus any method which measures a different ratio of these distances will inevitably measure a different combination of convergence terms \cite{Birrer_2020,Birrer_2022}.

As pointed out in \cite{Teodori_2022}, this procedure represents a simplified version of the real procedure used to constrain $\kappa_\mathrm{ext}$ via the velocity dispersion, in which uncertainties from the inference of parameters such as $\gamma$ must be taken into account, as well as the role of the point spread and luminosity weighting functions, and projection effects must also be considered. Nonetheless, it demonstrates that the convergence terms affecting the velocity dispersion and lensing observables are not straightforwardly factored out.
    \subsection{Estimating the bias}
    \label{sec:H0_bias}

\begin{figure}
    \centering
    \includegraphics[]{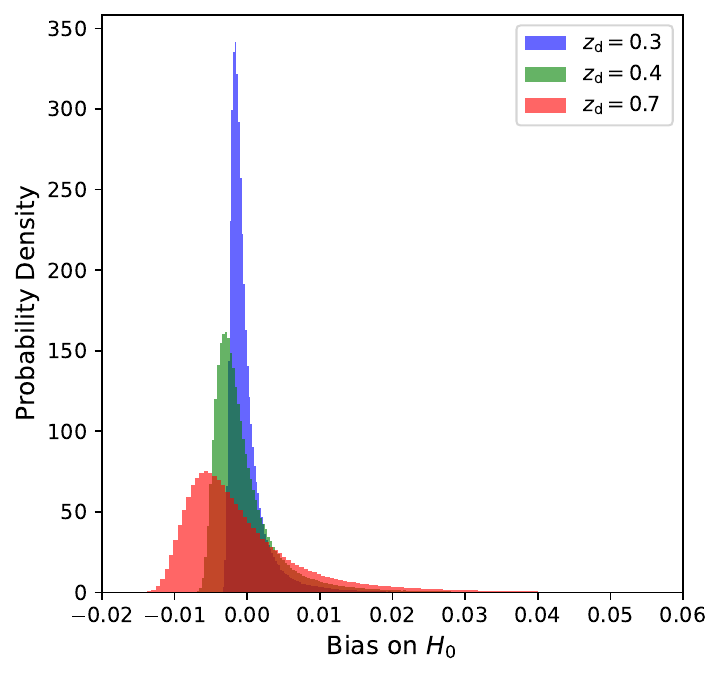}
    \caption{The probability density function for $H_0^\mathrm{inferred}/H_0-1$ at different redshifts. The standard deviation for $z=0.3$ is $\sim 0.4\%$, for $z=0.45$ is $\sim 0.7\%$ and for $z=0.7$ is $\sim 1.1\%$. All distributions have a mean value of 0.}
    \label{fig:histograms}
\end{figure}
Figure \cref{fig:histograms} shows distributions of the foreground convergence $\kappa_\od$ at different redshifts, obtained as in \cref{sec:ellipticity_bias}. The distributions in \cref{fig:histograms} are not symmetric, as a minimum value of the convergence is set by what would be an empty line of sight, while no such sharp upper bound exists. Nonetheless, these distributions are centered on 0 for lines of sight drawn randomly, and so we would expect that $H_0$ values measured from strong lensing systems distributed randomly in space would have an additional scatter but no systematic bias. If, however, these systems are preferentially located in lines of sight with overdense foregrounds, these measurements would not only have a higher scatter, but $H_0$ might additionally be systematically overestimated. It is clear from the lens equation that lenses with overdense foregrounds would indeed be preferentially selected, as would those with underdense backgrounds, due to the appearance of the $(1-\kappa_\ds)^{-1}$ term in $(1-\kappa_\mathrm{LOS})$. Nonetheless, it is only the foreground which is relevant for our consideration, as, assuming $\kappa_\od$ and $\kappa_\ds$ are generally uncorrelated, selection effects relating to $\kappa_\ds$ should be accounted for in lens kinematics, while the $\kappa_\od$ bias remains.

Lensing galaxies (typically early-type) are more likely to be found in over-dense regions of space \cite{Keeton_1997}, in which the typical convergence and shear are higher. Cosmological simulations demonstrate that the lensing efficiency of clusters can be significantly increased by line-of-sight structure \cite{Bartelmann_1998,Meneghetti_2013,Puchwein_2009}, which would once again suggest that observed strong lenses are affected by above-average external convergences. This is further supported by observations in \cite{Bayliss_2014}, which revealed a pronounced overdensity of galaxies along the line of sight to strong lensing galaxy clusters, and \cite{wells-2024}, which found that the mean external convergence of strong lensing systems was greater than zero, and increased with increasing source and deflector redshift. If these overdensities consist only of matter associated with the main lens galaxy, then we would expect these to impact the $\kappa_\mathrm{LOS}$ term and not $\kappa_\od$, as the former has a lensing efficiency which peaks closer to the main lens, where the latter's goes to zero (see e.g. \cite{Fleury_2021a}). This was the conclusion of \cite{Wong_2019}, who found that, for the sample of 87 strong lenses investigated, the line-of-sight overdensity associated with lensing systems arose only from the environments of the lenses themselves. However, \cite{Puchwein_2009,Bayliss_2014} in particular find clear evidence that the lensing efficiency increases with large-scale structure along the line of sight which is uncorrelated with the main lens galaxy, which would suggest that there could indeed be a systematic bias towards overestimations of $H_0$. 

\section{Conclusion}
\label{sec:conclusion}

In this work, we have explored degeneracies between the main lens and the foreground line of sight, and their consequences for the interpretation of lensing observables. While much of the qualitative discussion can be found in the existing literature, we have added explicit analytical relationships in \cref{eq:azimuthal_degeneracy,eq:degeneracy_form,eq:radial_observable2,eq:gamma}, and quantified the biases on lensing observables in \cref{sec:ellipticity_bias,sec:H0_bias}. We have argued that the main lens' ellipticity and the foreground shear are degenerate with one another, and derived the explicit form of this degeneracy for the popular elliptical power law lens model. This degeneracy prevents reliable measurements of the foreground shear, and adds an additional uncertainty of a few percent into measurements of the ellipticity of the lens mass distribution, and of a few degrees when comparing the lens mass and lens light profile inclinations. We have also shown that, for the EPL, this degeneracy affects only the angular structure of the lens, and measurements of the Einstein radius of the lens should be unbiased with respect to this foreground shear. This is, however, not the case for the foreground convergence, and attempts to constrain the mass-sheet degeneracy with velocity dispersion measurements can introduce an additional uncertainty on the order of a percent into $H_0$ measurements. Estimating the size of this uncertainty is limited by the resolution of current cosmological simulations, and larger variability in the foreground convergence within the angular scales of typical lensing systems could increase this further. For random lines of sight, $H_0$ measurements should nonetheless remain unbiased, but if strong lenses are located preferentially in lines of sight which are overdense as a result of matter which is uncorrelated with the lens galaxy or cluster, as several studies have suggested, we would expect $H_0$ measurements from time-delay cosmography to be systematically overestimated. A full quantitative analysis of these selection effects will be the topic of future research.

Foreground weak lensing effects are far from the only contribution to the error budget in time-delay cosmography, and it is crucial that uncertainties in measurements of relative time delays, assumptions on the stellar anisotropy profile, measurements of the lens profile slope, microlensing effects etc are well understood and well constrained before strong lensing is cemented as a reliable and competitive cosmological probe. Nonetheless, we emphasise that the foreground effects described in this paper are first order, and the foreground convergence and shear should form part of a self-consistent model of line-of-sight effects when trying to constrain $H_0$ or the angular properties of the lensing galaxy respectively. 

Despite these challenges, continuous progress is being made towards identifying and eliminating systematics in time-delay cosmography, while increasingly sophisticated mass and velocity dispersion and observational techniques are being employed, and there is good reason for optimism about the future of gravitational lensing as an observational tool. 

\section*{Acknowledgements}

DJ thanks Simon Birrer, Sherry Suyu, Anowar Shajib, Simon Dye and Thomas Collett for their discussion during the Multi-Messenger Gravitational Lensing and Strong Gravitational Lensing Science workshops in Manchester and Oxford. This work was supported by the First Rand Foundation, the Centre National de la Recherche Scientifique of France, David \& Elaine Potter Foundation and the National Astrophysics and Space Science Program of South Africa (NASSP). PF acknowledges support from the French \emph{Agence Nationale de la Recherche} through the ELROND project (ANR-23-CE31-0002). LM acknowledges financial support from the Inter-University Institute for Data Intensive Astronomy (IDIA), a partnership of the University of Cape Town, the University of Pretoria and the University of the Western Cape, and from the South African Department of Science and Innovation’s National Research Foundation under the ISARP RADIOMAP Joint Research Scheme (DSI-NRF Grant Number 150551) and the CPRR Projects (DSI-NRF Grant Number SRUG2204254729).

\section*{\href{https://www.elsevier.com/authors/policies-and-guidelines/credit-author-statement}{CRedIT} authorship contribution statement}

\noindent \textbf{Daniel Johnson:} Conceptualisation, Methodology, Formal Analysis, Writing - Original draft. 
\textbf{Pierre Fleury:} Conceptualisation, Validation, Writing - Review \& Editing, Visualization, Supervision, Project administration.
\textbf{Julien Larena:} Conceptualisation, Validation, Writing - Review \& Editing, Supervision.
\textbf{Lucia Marchetti:} Supervision, Writing - Review \& Editing, Project administration.

\appendix

\section{The mass-sheet degeneracy}
\label{appendix}
%
\subsection{The unobservability of the line-of-sight convergence}
\label{sec:MST}

Suppose each amplification matrix was transformed according to 
\begin{align}
    \bA_{ij}^\lambda = (1-\lambda_{ij})\bA_{ij}.
\end{align}
Parameterised as normal in terms of convergence and shear, the resulting matrices would take the form
\begin{equation}
    \bA^\lambda_{ij} = \begin{pmatrix} 1-\lambda_{ij}-(1-\lambda_{ij})\kappa_{ij}-(1-\lambda_{ij})\gamma^1_{ij}& -(1-\lambda_{ij})\gamma^2_{ij} \\ -(1-\lambda_{ij})\gamma^2_{ij} & 1-\lambda_{ij} - (1-\lambda_{ij})\kappa_{ij} + (1-\lambda_{ij})\gamma^1_{ij} \end{pmatrix}.
\end{equation}
From this result, we can read off
\begin{align}
    \kappa^\lambda_{ij} &= (1-\lambda_{ij})\kappa_{ij}+\lambda_{ij},  \\
    \gamma^\lambda_{ij} &= (1-\lambda_{ij})\gamma_{ij},
    \label{eq:los_convergence_shear_msd}
\end{align}
Now, consider the minimal lens model. Under the above transformation, this becomes
\begin{align}
    \tilde{\bbeta}^\lambda &= \ALOS^\lambda\btheta-\frac{\rd \psi^\lambda_{{\rm eff}}}{\rd \btheta},  \\
    \frac{1-\lambda_{\od}}{1-\lambda_{\ds}}\tilde{\bbeta} &= \frac{(1-\lambda_{\od})(1-\lambda_{\os})}{1-\lambda_{\ds}}\ALOS\btheta-\frac{\rd \psi_{{\rm eff}}[(1-\lambda_{\od})\btheta]}{\rd \btheta}.
\end{align}
Multiplying throughout by $(1-\lambda_{\ds})(1-\lambda_{\od})^{-1}(1-\lambda_{\os})^{-1}$, 
\begin{align}
    \frac{1}{1-\lambda_{\os}}\tilde{\bbeta} &= \ALOS\btheta-\frac{1-\lambda_{\ds}}{(1-\lambda_{\od})(1-\lambda_{\os})}\frac{\rd \psi_{{\rm eff}}[(1-\lambda_{\od})\btheta]}{\rd \btheta},
\end{align}
or equivalently, 
\begin{equation}
    \tilde{\bbeta}' = \ALOS\btheta-\frac{\rd \psi_{{\rm eff}}'}{\rd \btheta},
\end{equation}
where
\begin{gather}
    \tilde{\bbeta}' \equiv \frac{1}{1-\lambda_{\os}}\bA_{\od}\bA_{\ds}^{-1}\bbeta,  \\
    \psi_{{\rm eff}}'(\btheta) \equiv \frac{1-\lambda_{\ds}}{(1-\lambda_{\od})(1-\lambda_{\os})}\psi_{{\rm eff}}[(1-\lambda_{\od})\btheta].
\end{gather}
As the source position is unobservable, it is clear that $\tilde{\bbeta}'$ is degenerate with $\tilde{\bbeta}$ and $\bbeta$. The $(1-\lambda_{\ds})(1-\lambda_{\od})^{-1}(1-\lambda_{\os})^{-1}$ term which multiplies $\psi_{{\rm eff}}$ can be fully absorbed into the lens model itself by uniformly rescaling its potential by the same factor. The $(1-\lambda_{\od})$ term within the argument of $\psi_{{\rm eff}}$ is also degenerate with the properties of the main lens. From this and \cref{eq:los_convergence_shear_msd}, it is clear that, without additional information about the lens profile, the line-of-sight convergences are degenerate with properties of the source and the main lens. The line-of-sight shears are also scaled by the unobservable $(1-\lambda_{ij})$ factors, and only the reduced shear terms are unaffected by this degeneracy.

\subsection{Effects on $H_0$}

Under the MST described in \cref{sec:MST} and using \cref{eq:lambda_los}, the new time delay function \cref{eq:time_delay_tidal_general} will be
\begin{align}
    t^\lambda(\btheta,\bbeta) =& \tau_{\ds}\left[\frac{1}{2}(\btheta-\bbeta'_\lambda)\cdot\bA^\lambda_{{\rm LOS}}(\btheta-\bbeta'_\lambda)-\psi^\lambda_{{\rm eff}}(\btheta) \right], \\
    =& \tau_{\ds}\left\{\frac{1}{2}\left(\btheta-\frac{1}{1-\lambda_{\os}}\bbeta'\right)\cdot(1-\lambda_\mathrm{LOS})\bA_{{\rm LOS}}\left(\btheta-\frac{1}{1-\lambda_{\os}}\bbeta'\right)-\psi_{{\rm eff}}\left[(1-\lambda_{\od})\btheta\right] \right\}, \\
    =& (1-\lambda_\mathrm{LOS})\tau_{\ds}\left[\frac{1}{2}\left(\btheta-\frac{1}{1-\lambda_{\os}}\bbeta'\right)\cdot\bA_{{\rm LOS}}\left(\btheta-\frac{1}{1-\lambda_{\os}}\bbeta'\right)-\psi'_{{\rm eff}}(\btheta) \right].
    \label{eq:mst_td}
\end{align}
Because the $(1-\lambda_\mathrm{LOS})$ term is independent of the specific image $\btheta_i$, and because the source position is unobservable, a measured relative time delay of $\Delta t$ is proportional to both $(1-\lambda_\mathrm{LOS})$ and $\tau_{\ds}$, and the effects of these terms are indistinguishable. Notice also that, while the source terms appearing in \cref{eq:mst_td,eq:minimal_lens} are different, the factor $1-\lambda_{\os}$ appears identically in both, and thus this cannot break the degeneracy.

Expanding this prefactor in terms of the deflector redshift $z_\mathrm{d}$ and angular diameter distances, we have that
\begin{equation}
    (1-\lambda_\mathrm{LOS})\tau_{\ds} = (1-\lambda_\mathrm{LOS})(1+z_\mathrm{d})\frac{D_\od D_\os}{cD_\ds}.
\end{equation}
Each of the angular diameter distances is proportional to $H_0^{-1}$, and so
\begin{align}
    {\Delta t} &\propto (1-\lambda_\mathrm{LOS})\tau_{\ds}, \\
    {\Delta t} &\propto (1-\lambda_\mathrm{LOS})\frac{D_\od D_\os}{cD_\ds}, \\
    H_0  &\propto \frac{1-\lambda_\mathrm{LOS}}{\Delta t}.
\end{align}
The true value of $H_0$ can be found if one knows the true value of $\Delta t$ that one measures, and the true value of the MST factor. From the way in which we have defined $\lambda_\mathrm{LOS}$, this term corresponds to a measurement bias and should be zero. With a non-zero MST factor, then the measured $H_0^{\mathrm{model}}$ will be biased by a factor $(1-\lambda_\mathrm{LOS})$, and the true value of $H_0$ will relate to this via
\begin{equation}
    H_0 = \frac{1}{1-\lambda_\mathrm{LOS}}H_0^\mathrm{model}.
\end{equation}
The typical choice is to ignore convergences in the lens model, which, from \cref{eq:los_convergence_shear_msd} would be equivalent to setting the MST factor to
\begin{align}
    \kappa^\lambda_\mathrm{LOS} &= 0, \\
    (1-\lambda_\mathrm{LOS})\kappa_\mathrm{LOS}+\lambda_\mathrm{LOS} &= 0, \\
    1-\lambda_\mathrm{LOS} &= \frac{1}{1-\kappa_\mathrm{LOS}},
\end{align}
from which \cref{eq:true_correction} follows.\\
%
\bibliographystyle{JHEP.bst}
\bibliography{Bibliography.bib}

\end{document}

%% file: packages.tex
\usepackage{jcappub}

\usepackage{mathrsfs}
\usepackage{bbm}
\usepackage{bm}
\usepackage{dsfont}
\usepackage{tensor}
\usepackage{xspace}
\usepackage{lmodern}
\usepackage{wasysym}
\usepackage{microtype}
\usepackage{empheq}
\usepackage{ifthen}
\usepackage{amsmath}
\usepackage{wrapfig}
\usepackage{cleveref}
\usepackage{xcolor}
\usepackage{siunitx}
\DeclareSIUnit\parsec{pc}
\usepackage{floatrow}
\usepackage[normalem]{ulem}

%% file: macros.tex
\newcommand\rd{{\rm d}}
\newcommand\balpha{\bm{\alpha}}
\newcommand\btheta{\bm{\theta}}

\newcommand\bG{\bm{G}}
\newcommand\bbeta{\bm{\beta}}

\newcommand\ualpha{\underline{\alpha}}
\newcommand\utheta{\underline{\theta}}

\newcommand\ri{{\rm i}}
\newcommand\re{{\rm e}}

\newcommand\bA{\bm{\mathcal{A}}}

\newcommand\ALOS{\bA_{{\rm LOS}}}

\newcommand\ext{{\rm ext}}
\newcommand\Sersic{S\'ersic }

\newcommand\thetaE{\theta_{\mathrm{E}}}
\newcommand\tthetaE{\tilde{\theta}_{\mathrm{E}}}
\newcommand\os{\mathrm{os}}
\newcommand\ds{\mathrm{ds}}
\newcommand\od{\mathrm{od}}

\newcommand{\delimiters}[4][]{
\ifthenelse{ \equal{#1}{1} }{  #2 #3 #4  }
					{ \ifthenelse{\equal{#1}{2}}{ \big#2 #3 \big#4 }
						{ \ifthenelse{\equal{#1}{3}}{ \Big#2 #3 \Big#4 }
							{ \ifthenelse{\equal{#1}{4}}{ \bigg#2 #3 \bigg#4 }
								{ \ifthenelse{\equal{#1}{5}}{ \Bigg#2 #3 \Bigg#4 }
									{ \left#2 #3 \right#4 }
								}
							}
						}
					}
													}

\definecolor{blue4}{RGB}{0,0,143}
\definecolor{red4}{RGB}{143,0,0}
\definecolor{orange}{RGB}{255,128,0}
\definecolor{darkcyan}{RGB}{0,128,128}
\definecolor{olive}{RGB}{0,128,0}
\definecolor{purple}{RGB}{128,0,128}
\definecolor{cyan2}{RGB}{0,255,255}
\definecolor{fushia}{RGB}{255,0,255}
\definecolor{mygray}{gray}{0.5}
\definecolor{lightgray}{gray}{0.85}


%% file: main.bbl
\providecommand{\href}[2]{#2}\begingroup\raggedright\begin{thebibliography}{10}

\bibitem{Abbot_2018}
{DES and SPT Collaborations}, T.~M.~C. Abbott, et~al., {\it {Dark Energy Survey Year 1 Results: A Precise {H}0 Estimate from DES Y1, BAO, and D/H Data}},  {\em Monthly Notices of the Royal Astronomical Society} {\bf 480} (07, 2018) 3879--3888, [\href{http://arxiv.org/abs/1711.00403}{{\tt arXiv:1711.00403}}].

\bibitem{Cuceu_2019}
A.~Cuceu, J.~Farr, P.~Lemos, and A.~Font-Ribera, {\it Baryon acoustic oscillations and the hubble constant: past, present and future},  {\em Journal of Cosmology and Astroparticle Physics} {\bf 2019} (Oct., 2019) 044–044, [\href{http://arxiv.org/abs/2112.04510}{{\tt arXiv:2112.04510}}].

\bibitem{Planck_2020}
{Planck Collaboration}, N.~{Aghanim}, et~al., {\it {Planck 2018 results. VI. Cosmological parameters}},  {\em Astronomy \& Astrophysics} {\bf 641} (Sept., 2020) A6, [\href{http://arxiv.org/abs/1807.06209}{{\tt arXiv:1807.06209}}].

\bibitem{Riess_2022}
A.~G. Riess, W.~Yuan, L.~M. Macri, D.~Scolnic, D.~Brout, S.~Casertano, D.~O. Jones, Y.~Murakami, G.~S. Anand, L.~Breuval, T.~G. Brink, A.~V. Filippenko, S.~Hoffmann, S.~W. Jha, W.~D’arcy~Kenworthy, J.~Mackenty, B.~E. Stahl, and W.~Zheng, {\it A comprehensive measurement of the local value of the {H}ubble {C}onstant with $1~\mathrm{km\,s^{-1}\,Mpc^{-1}}$ uncertainty from the {H}ubble {S}pace {T}elescope and the {S}{H}0{E}{S} team},  {\em The Astrophysical Journal Letters} {\bf 934} (July, 2022) L7, [\href{http://arxiv.org/abs/2112.04510}{{\tt arXiv:2112.04510}}].

\bibitem{Di_Valentino_2021}
E.~D. Valentino, O.~Mena, S.~Pan, L.~Visinelli, W.~Yang, A.~Melchiorri, D.~F. Mota, A.~G. Riess, and J.~Silk, {\it In the realm of the {H}ubble tension—a review of solutions*},  {\em Classical and Quantum Gravity} {\bf 38} (07, 2021) 153001, [\href{http://arxiv.org/abs/2103.01183}{{\tt arXiv:2103.01183}}].

\bibitem{Freedman_2021}
W.~L. Freedman, {\it Measurements of the {H}ubble {C}onstant: {T}ensions in {P}erspective},  {\em The Astrophysical Journal} {\bf 919} (sep, 2021) 16, [\href{http://arxiv.org/abs/2106.15656}{{\tt arXiv:2106.15656}}].

\bibitem{Abdalla_2022}
E.~{Abdalla} et~al., {\it {Cosmology intertwined: A review of the particle physics, astrophysics, and cosmology associated with the cosmological tensions and anomalies}},  {\em Journal of High Energy Astrophysics} {\bf 34} (June, 2022) 49--211, [\href{http://arxiv.org/abs/2203.06142}{{\tt arXiv:2203.06142}}].

\bibitem{Keeton_1997}
C.~R. Keeton, C.~S. Kochanek, and U.~Seljak, {\it Shear and ellipticity in gravitational lenses},  {\em The Astrophysical Journal} {\bf 482} (06, 1997) 604--620, [\href{http://arxiv.org/abs/astro-ph/9610163}{{\tt astro-ph/9610163}}].

\bibitem{Oguri_2007}
M.~Oguri, {\it Gravitational lens time delays: A statistical assessment of lens model dependences and implications for the global {H}ubble {C}onstant},  {\em The Astrophysical Journal} {\bf 660} (May, 2007) 1–15, [\href{http://arxiv.org/abs/astro-ph/0609694}{{\tt astro-ph/0609694}}].

\bibitem{Suyu_2010}
S.~H. Suyu, P.~J. Marshall, M.~W. Auger, S.~Hilbert, R.~D. Blandford, L.~V.~E. Koopmans, C.~D. Fassnacht, and T.~Treu, {\it Dissecting the gravitational lens {B1608+656}. {II}. precision measurements of the {H}ubble {C}onstant, spatial curvature, and the dark energy equation of state},  {\em The Astrophysical Journal} {\bf 711} (02, 2010) 201, [\href{http://arxiv.org/abs/0910.2773}{{\tt arXiv:0910.2773}}].

\bibitem{Yildirim_2021}
A.~Yıldırım, S.~H. Suyu, G.~C.~F. Chen, and E.~Komatsu, {\it {TDCOSMO} {VIII}: Cosmological distance measurements in light of the mass-sheet degeneracy -- forecasts from strong lensing and integrated field unit stellar kinematics},  {\em Astronomy \& Astrophysics} {\bf 675} (2023) A21, [\href{http://arxiv.org/abs/2109.14615}{{\tt arXiv:2109.14615}}].

\bibitem{Birrer_2022}
S.~Birrer, M.~Millon, D.~Sluse, A.~J. Shajib, F.~Courbin, L.~V.~E. Koopmans, S.~H. Suyu, and T.~Treu, {\it Time-delay cosmography: Measuring the {H}ubble {C}onstant and other cosmological parameters with strong gravitational lensing},  \href{http://arxiv.org/abs/2210.10833}{{\tt arXiv:2210.10833}}.

\bibitem{Treu_2022}
T.~Treu, S.~H. Suyu, and P.~J. Marshall, {\it Strong lensing time-delay cosmography in the 2020s},  {\em The Astronomy and Astrophysics Review} {\bf 30} (nov, 2022) [\href{http://arxiv.org/abs/2210.15794}{{\tt arXiv:2210.15794}}].

\bibitem{Refsdal_1964}
S.~Refsdal, {\it On the possibility of determining {H}ubble's parameter and the masses of galaxies from the gravitational lens effect},  {\em Monthly Notices of the Royal Astronomical Society} {\bf 128} (Jan., 1964) 307, [\href{http://arxiv.org/abs/https://academic.oup.com/mnras/article/128/4/307/2601707}{{\tt https://academic.oup.com/mnras/article/128/4/307/2601707}}].

\bibitem{Refsdal_1966}
S.~Refsdal and S.~Rosseland, {\it {On the Possibility of Determining the Distances and Masses of Stars from the Gravitational Lens Effect}},  {\em Monthly Notices of the Royal Astronomical Society} {\bf 134} (12, 1966) 315--319, [\href{http://arxiv.org/abs/https://academic.oup.com/mnras/article-pdf/134/3/315/8075171/mnras134-0315.pdf}{{\tt https://academic.oup.com/mnras/article-pdf/134/3/315/8075171/mnras134-0315.pdf}}].

\bibitem{Bonvin_2016}
V.~Bonvin, F.~Courbin, S.~H. Suyu, P.~J. Marshall, C.~E. Rusu, D.~Sluse, M.~Tewes, K.~C. Wong, T.~Collett, C.~D. Fassnacht, T.~Treu, M.~W. Auger, S.~Hilbert, L.~V.~E. Koopmans, G.~Meylan, N.~Rumbaugh, A.~Sonnenfeld, and C.~Spiniello, {\it H0{L}i{COW} {V}. new {COSMOGRAIL} time delays of {HE}~0435-1223: ${H}_0$ to 3.8~percent precision from strong lensing in a flat $\lambda${CDM} model},  {\em Monthly Notices of the Royal Astronomical Society} {\bf 465} (11, 2016) 4914--4930, [\href{http://arxiv.org/abs/1607.01790}{{\tt arXiv:1607.01790}}].

\bibitem{Millon_2020}
M.~Millon, F.~Courbin, V.~Bonvin, E.~Buckley-Geer, C.~D. Fassnacht, J.~Frieman, P.~J. Marshall, S.~H. Suyu, T.~Treu, T.~Anguita, V.~Motta, A.~Agnello, J.~H.~H. Chan, D.~C.-Y. Chao, M.~Chijani, D.~Gilman, K.~Gilmore, C.~Lemon, J.~R. Lucey, A.~Melo, E.~Paic, K.~Rojas, D.~Sluse, P.~R. Williams, A.~Hempel, S.~Kim, R.~Lachaume, and M.~Rabus, {\it {TDCOSMO} {I}. an exploration of systematic uncertainties in the inference of ${H}_0$ from time-delay cosmography},  {\em Astronomy {\&} Astrophysics} {\bf 642} (10, 2020) A193, [\href{http://arxiv.org/abs/1912.08027}{{\tt arXiv:1912.08027}}].

\bibitem{Kumar_2015}
S.~R. Kumar, C.~S. Stalin, and T.~P. Prabhu, {\it ${H}_0$ from ten well-measured time delay lenses},  {\em Astronomy {\&} Astrophysics} {\bf 580} (07, 2015) A38, [\href{http://arxiv.org/abs/1404.2920}{{\tt arXiv:1404.2920}}].

\bibitem{Tewes_2013}
M.~Tewes, F.~Courbin, and G.~Meylan, {\it {COSMOGRAIL}: the {COSmological} {MOnitoring} of {GRAvItational} lenses {XI}. techniques for time delay measurement in presence of microlensing},  {\em Astronomy {\&} Astrophysics} {\bf 553} (05, 2013) A120, [\href{http://arxiv.org/abs/1208.5598}{{\tt arXiv:1208.5598}}].

\bibitem{Chen_2019}
G.~C.-F. Chen, C.~D. Fassnacht, S.~H. Suyu, C.~E. Rusu, J.~H.~H. Chan, K.~C. Wong, M.~W. Auger, S.~Hilbert, V.~Bonvin, S.~Birrer, M.~Millon, L.~V.~E. Koopmans, D.~J. Lagattuta, J.~P. McKean, S.~Vegetti, F.~Courbin, X.~Ding, A.~Halkola, I.~Jee, A.~J. Shajib, D.~Sluse, A.~Sonnenfeld, and T.~Treu, {\it A {SHARP} view of {H}0{L}i{COW}: {H}0 from three time-delay gravitational lens systems with adaptive optics imaging},  {\em Monthly Notices of the Royal Astronomical Society} {\bf 490} (09, 2019) 1743--1773, [\href{http://arxiv.org/abs/1907.02533}{{\tt arXiv:1907.02533}}].

\bibitem{Suyu_2013}
S.~H. Suyu, M.~W. Auger, S.~Hilbert, P.~J. Marshall, M.~Tewes, T.~Treu, C.~D. Fassnacht, L.~V.~E. Koopmans, D.~Sluse, R.~D. Blandford, F.~Courbin, and G.~Meylan, {\it Two accurate time-delay distances from strong lensing: implications for cosmology},  {\em The Astrophysical Journal} {\bf 766} (03, 2013) 70, [\href{http://arxiv.org/abs/1208.6010}{{\tt arXiv:1208.6010}}].

\bibitem{Gorenstein_1988}
M.~V. {Gorenstein}, E.~E. {Falco}, and I.~I. {Shapiro}, {\it {Degeneracies in Parameter Estimates for Models of Gravitational Lens Systems}},  {\em Astrophysical Journal} {\bf 327} (04, 1988) 693, [\href{http://arxiv.org/abs/https://ui.adsabs.harvard.edu/abs/1988ApJ...327..693G}{{\tt https://ui.adsabs.harvard.edu/abs/1988ApJ...327..693G}}].

\bibitem{Saha_2000}
P.~Saha, {\it Lensing degeneracies revisited},  {\em The Astronomical Journal} {\bf 120} (10, 2000) 1654--1659, [\href{http://arxiv.org/abs/astro-ph/0006432}{{\tt astro-ph/0006432}}].

\bibitem{Wucknitz_2002}
O.~Wucknitz, {\it Degeneracies and scaling relations in general power-law models for gravitational lenses},  {\em Monthly Notices of the Royal Astronomical Society} {\bf 332} (06, 2002) 951--961, [\href{http://arxiv.org/abs/astro-ph/0202376}{{\tt astro-ph/0202376}}].

\bibitem{Liesenborgs_2012}
J.~Liesenborgs and S.~De~Rijcke, {\it Lensing degeneracies and mass substructure},  {\em Monthly Notices of the Royal Astronomical Society} {\bf 425} (2012), no.~3 1772--1780, [\href{http://arxiv.org/abs/1207.4692}{{\tt arXiv:1207.4692}}].

\bibitem{Falco_1985}
E.~E. {Falco}, M.~V. {Gorenstein}, and I.~I. {Shapiro}, {\it {On model-dependent bounds on {H}0 from gravitational images : application to Q 0957+561 A, B.}},  {\em Astrophysical Journal, Letters} {\bf 289} (02, 1985) L1--L4, [\href{http://arxiv.org/abs/https://ui.adsabs.harvard.edu/abs/1985ApJ...289L...1F}{{\tt https://ui.adsabs.harvard.edu/abs/1985ApJ...289L...1F}}].

\bibitem{Schneider_2013}
P.~Schneider and D.~Sluse, {\it Mass-sheet degeneracy, power-law models and external convergence: Impact on the determination of the hubble constant from gravitational lensing},  {\em Astronomy {\&} Astrophysics} {\bf 559} (11, 2013) A37, [\href{http://arxiv.org/abs/1306.0901}{{\tt arXiv:1306.0901}}].

\bibitem{Keeton_2004}
C.~R. Keeton and A.~I. Zabludoff, {\it The importance of lens galaxy environments},  {\em The Astrophysical Journal} {\bf 612} (Sept., 2004) 660–678, [\href{http://arxiv.org/abs/astro-ph/0406060}{{\tt astro-ph/0406060}}].

\bibitem{Fassnacht_2011}
C.~D. Fassnacht, L.~V.~E. Koopmans, and K.~C. Wong, {\it {Galaxy number counts and implications for strong lensing}},  {\em Monthly Notices of the Royal Astronomical Society} {\bf 410} (01, 2011) 2167--2179, [\href{http://arxiv.org/abs/0909.4301}{{\tt arXiv:0909.4301}}].

\bibitem{Kochanek_2020}
C.~S. Kochanek, {\it {Overconstrained gravitational lens models and the Hubble constant}},  {\em Monthly Notices of the Royal Astronomical Society} {\bf 493} (02, 2020) 1725--1735, [\href{http://arxiv.org/abs/1911.05083}{{\tt arXiv:1911.05083}}].

\bibitem{Fassnacht_2006}
C.~D. Fassnacht, R.~R. Gal, L.~M. Lubin, J.~P. McKean, G.~K. Squires, and A.~C.~S. Readhead, {\it Mass along the line of sight to the gravitational lens {B}1608+656: Galaxy groups and implications for ${H}_0$},  {\em The Astrophysical Journal} {\bf 642} (05, 2006) 30, [\href{http://arxiv.org/abs/astro-ph/0510728}{{\tt astro-ph/0510728}}].

\bibitem{Momcheva_2006}
I.~Momcheva, K.~Williams, C.~Keeton, and A.~Zabludoff, {\it A spectroscopic study of the environments of gravitational lens galaxies},  {\em The Astrophysical Journal} {\bf 641} (04, 2006) 169, [\href{http://arxiv.org/abs/astro-ph/0511594}{{\tt astro-ph/0511594}}].

\bibitem{Williams_2006}
K.~A. Williams, I.~Momcheva, C.~R. Keeton, A.~I. Zabludoff, and J.~Lehar, {\it First results from a photometric survey of strong gravitational lens environments},  {\em The Astrophysical Journal} {\bf 646} (July, 2006) 85–106, [\href{http://arxiv.org/abs/astro-ph/0511593}{{\tt astro-ph/0511593}}].

\bibitem{Auger_2007}
M.~W. Auger, C.~D. Fassnacht, A.~L. Abrahamse, L.~M. Lubin, and G.~K. Squires, {\it The gravitational lens-galaxy group connection. {II}. groups associated with {B}2319+051 and {B}1600+434},  {\em The Astronomical Journal} {\bf 134} (06, 2007) 668, [\href{http://arxiv.org/abs/astro-ph/0603448}{{\tt astro-ph/0603448}}].

\bibitem{Wilson_2017}
M.~L. Wilson, A.~I. Zabludoff, C.~R. Keeton, K.~C. Wong, K.~A. Williams, K.~D. French, and I.~G. Momcheva, {\it A spectroscopic survey of the fields of 28 strong gravitational lenses: Implications for ${H}_0$},  {\em The Astrophysical Journal} {\bf 850} (11, 2017) 94, [\href{http://arxiv.org/abs/1710.09900}{{\tt arXiv:1710.09900}}].

\bibitem{Springel_2005}
V.~{Springel}, S.~D.~M. {White}, A.~{Jenkins}, C.~S. {Frenk}, N.~{Yoshida}, L.~{Gao}, J.~{Navarro}, R.~{Thacker}, D.~{Croton}, J.~{Helly}, J.~A. {Peacock}, S.~{Cole}, P.~{Thomas}, H.~{Couchman}, A.~{Evrard}, J.~{Colberg}, and F.~{Pearce}, {\it {Simulations of the formation, evolution and clustering of galaxies and quasars}},  {\em Nature} {\bf 435} (06, 2005) 629--636, [\href{http://arxiv.org/abs/astro-ph/0504097}{{\tt astro-ph/0504097}}].

\bibitem{Greene_2013}
Z.~S. Greene, S.~H. Suyu, T.~Treu, S.~Hilbert, M.~W. Auger, T.~E. Collett, P.~J. Marshall, C.~D. Fassnacht, R.~D. Blandford, M.~Bradač, and L.~V.~E. Koopmans, {\it Improving the precision of time-delay cosmography with observations of galaxies along the line of sight},  {\em The Astrophysical Journal} {\bf 768} (04, 2013) 39, [\href{http://arxiv.org/abs/1303.3588}{{\tt arXiv:1303.3588}}].

\bibitem{Birrer_2017}
S.~Birrer, C.~Welschen, A.~Amara, and A.~Refregier, {\it Line-of-sight effects in strong lensing: putting theory into practice},  {\em Journal of Cosmology and Astroparticle Physics} {\bf 2017} (04, 2017) 049--049, [\href{http://arxiv.org/abs/1610.01599}{{\tt arXiv:1610.01599}}].

\bibitem{Fischer_1997}
P.~{Fischer}, G.~{Bernstein}, G.~{Rhee}, and J.~A. {Tyson}, {\it {The Mass Distribution of the Cluster 0957+561 From Gravitational Lensing}},  {\em The Astrophysical Journal} {\bf 113} (02, 1997) 521, [\href{http://arxiv.org/abs/astro-ph/9608117}{{\tt astro-ph/9608117}}].

\bibitem{Nakajima_2009}
R.~Nakajima, G.~M. Bernstein, R.~Fadely, C.~R. Keeton, and T.~Schrabback, {\it Improved constraints on the gravitational lens {Q}0957+561. {I}. weak lensing},  {\em The Astrophysical Journal} {\bf 697} (05, 2009) 1793, [\href{http://arxiv.org/abs/0903.4182}{{\tt arXiv:0903.4182}}].

\bibitem{Fadely_2010}
R.~Fadely, C.~R. Keeton, R.~Nakajima, and G.~M. Bernstein, {\it Improved constraints on the gravitational lens {Q}0957+561. {II}. strong lensing},  {\em The Astrophysical Journal} {\bf 711} (02, 2010) 246, [\href{http://arxiv.org/abs/0909.1807}{{\tt arXiv:0909.1807}}].

\bibitem{Tihhonova_2018}
O.~Tihhonova, F.~Courbin, D.~Harvey, S.~Hilbert, C.~E. Rusu, C.~D. Fassnacht, V.~Bonvin, P.~J. Marshall, G.~Meylan, D.~Sluse, S.~H. Suyu, T.~Treu, and K.~C. Wong, {\it {{H}0{L}i{COW} {VIII}. A weak-lensing measurement of the external convergence in the field of the lensed quasar {HE} 0435-1223}},  {\em Monthly Notices of the Royal Astronomical Society} {\bf 477} (04, 2018) 5657--5669, [\href{http://arxiv.org/abs/1711.08804}{{\tt arXiv:1711.08804}}].

\bibitem{Fleury_2021a}
P.~Fleury, J.~Larena, and J.-P. Uzan, {\it {Line-of-sight effects in strong gravitational lensing}},  {\em {JCAP}} {\bf 08} (2021) 024, [\href{http://arxiv.org/abs/2104.08883}{{\tt arXiv:2104.08883}}].

\bibitem{McCully_2017}
C.~McCully, C.~R. Keeton, K.~C. Wong, and A.~I. Zabludoff, {\it Quantifying environmental and line-of-sight effects in models of strong gravitational lens systems},  {\em The Astrophysical Journal} {\bf 836} (02, 2017) 141, [\href{http://arxiv.org/abs/1601.05417}{{\tt arXiv:1601.05417}}].

\bibitem{McCully_2014}
C.~McCully, C.~R. Keeton, K.~C. Wong, and A.~I. Zabludoff, {\it A new hybrid framework to efficiently model lines of sight to gravitational lenses},  {\em Monthly Notices of the Royal Astronomical Society} {\bf 443} (08, 2014) 3631--3642, [\href{http://arxiv.org/abs/1401.0197}{{\tt arXiv:1401.0197}}].

\bibitem{Rusu_2019}
C.~E. Rusu, K.~C. Wong, V.~Bonvin, D.~Sluse, S.~H. Suyu, C.~D. Fassnacht, J.~H.~H. Chan, S.~Hilbert, M.~W. Auger, A.~Sonnenfeld, S.~Birrer, F.~Courbin, T.~Treu, G.~C.-F. Chen, A.~Halkola, L.~V.~E. Koopmans, P.~J. Marshall, and A.~J. Shajib, {\it {H}0{L}i{COW} {XII}. lens mass model of {WFI}2033~-~4723~and blind measurement of its time-delay distance and ${H}_0$},  {\em Monthly Notices of the Royal Astronomical Society} {\bf 498} (09, 2019) 1440--1468, [\href{http://arxiv.org/abs/1905.09338}{{\tt arXiv:1905.09338}}].

\bibitem{Birrer_2019}
S.~Birrer, T.~Treu, C.~E. Rusu, V.~Bonvin, C.~D. Fassnacht, J.~H.~H. Chan, A.~Agnello, A.~J. Shajib, G.~C.-F. Chen, M.~Auger, F.~Courbin, S.~Hilbert, D.~Sluse, S.~H. Suyu, K.~C. Wong, P.~Marshall, B.~C. Lemaux, and G.~Meylan, {\it {H}0{L}i{COW} {\textendash} {IX}. cosmographic analysis of the doubly imaged quasar {SDSS} 1206$+$4332~and a new measurement of the {H}ubble constant},  {\em Monthly Notices of the Royal Astronomical Society} {\bf 484} (01, 2019) 4726--4753, [\href{http://arxiv.org/abs/1809.01274}{{\tt arXiv:1809.01274}}].

\bibitem{Wong_2019}
K.~C. Wong, S.~H. Suyu, G.~C.-F. Chen, C.~E. Rusu, M.~Millon, D.~Sluse, V.~Bonvin, C.~D. Fassnacht, S.~Taubenberger, M.~W. Auger, S.~Birrer, J.~H.~H. Chan, F.~Courbin, S.~Hilbert, O.~Tihhonova, T.~Treu, A.~Agnello, X.~Ding, I.~Jee, E.~Komatsu, A.~J. Shajib, A.~Sonnenfeld, R.~D. Blandford, L.~V.~E. Koopmans, P.~J. Marshall, and G.~Meylan, {\it {{H}0{L}i{COW} – XIII. A 2.4 per cent measurement of {H}0 from lensed quasars: 5.3$\sigma$ tension between early- and late-Universe probes}},  {\em Monthly Notices of the Royal Astronomical Society} {\bf 498} (09, 2019) 1420--1439, [\href{http://arxiv.org/abs/1907.04869}{{\tt arXiv:1907.04869}}].

\bibitem{Chen_2011}
J.~Chen, S.~M. Koushiappas, and A.~R. Zentner, {\it The effects of halo-to-halo variation on substructure lensing},  {\em The Astrophysical Journal} {\bf 741} (10, 2011) 117, [\href{http://arxiv.org/abs/1101.2916}{{\tt arXiv:1101.2916}}].

\bibitem{Shajib_2023}
A.~J. Shajib, P.~Mozumdar, G.~C.-F. Chen, T.~Treu, M.~Cappellari, S.~Knabel, S.~H. Suyu, V.~N. Bennert, J.~A. Frieman, D.~Sluse, S.~Birrer, F.~Courbin, C.~D. Fassnacht, L.~Villafaña, and P.~R. Williams, {\it {TDCOSMO}: {XII}. improved {H}ubble constant measurement from lensing time delays using spatially resolved stellar kinematics of the lens galaxy},  {\em Astronomy \& Astrophysics} {\bf 673} (Apr., 2023) A9, [\href{http://arxiv.org/abs/2301.02656}{{\tt arXiv:2301.02656}}].

\bibitem{Teodori_2022}
L.~Teodori, K.~Blum, E.~Castorina, M.~Simonovi{\'{c} }, and Y.~Soreq, {\it Comments on the mass sheet degeneracy in cosmography analyses},  {\em Journal of Cosmology and Astroparticle Physics} {\bf 2022} (07, 2022) 027, [\href{http://arxiv.org/abs/2201.05111}{{\tt arXiv:2201.05111}}].

\bibitem{Gomer_2021}
M.~R. Gomer and L.~L.~R. Williams, {\it Galaxy-lens determination of {H}0: the effect of the ellipse~$+$~shear modelling assumption},  {\em Monthly Notices of the Royal Astronomical Society} {\bf 504} (04, 2021) 1340--1354, [\href{http://arxiv.org/abs/1907.08638}{{\tt arXiv:1907.08638}}].

\bibitem{Etherington_2023}
A.~Etherington, J.~W. Nightingale, R.~Massey, S.-I. Tam, X.~Cao, A.~Niemiec, Q.~He, A.~Robertson, R.~Li, A.~Amvrosiadis, S.~Cole, J.~M. Diego, C.~S. Frenk, B.~L. Frye, D.~Harvey, M.~Jauzac, A.~M. Koekemoer, D.~J. Lagattuta, M.~Limousin, G.~Mahler, E.~Sirks, and C.~L. Steinhardt, {\it Strong gravitational lensing's `external shear' is not shear},  \href{http://arxiv.org/abs/2301.05244}{{\tt arXiv:2301.05244}}.

\bibitem{Hogg_2023}
N.~B. Hogg, P.~Fleury, J.~Larena, and M.~Martinelli, {\it Measuring line-of-sight shear with einstein rings: a proof of concept},  {\em Monthly Notices of the Royal Astronomical Society} {\bf 520} (Feb., 2023) 5982–6000, [\href{http://arxiv.org/abs/2210.07210}{{\tt arXiv:2210.07210}}].

\bibitem{Barkana_1998}
R.~Barkana, {\it Fast calculation of a family of elliptical gravitational lens models},  {\em The Astrophysical Journal} {\bf 502} (aug, 1998) 531, [\href{http://arxiv.org/abs/astro-ph/9802002}{{\tt astro-ph/9802002}}].

\bibitem{Tessore_2015}
N.~Tessore and R.~B. Metcalf, {\it The elliptical power law profile lens},  {\em Astronomy {\&} Astrophysics} {\bf 580} (08, 2015) A79, [\href{http://arxiv.org/abs/1507.01819}{{\tt arXiv:1507.01819}}].

\bibitem{Wong_2016}
K.~C. Wong, S.~H. Suyu, M.~W. Auger, V.~Bonvin, F.~Courbin, C.~D. Fassnacht, A.~Halkola, C.~E. Rusu, D.~Sluse, A.~Sonnenfeld, T.~Treu, T.~E. Collett, S.~Hilbert, L.~V.~E. Koopmans, P.~J. Marshall, and N.~Rumbaugh, {\it {H}0{L}i{COW} {\textendash} {IV}. lens mass model of {HE}~0435-1223 and blind measurement of its time-delay distance for cosmology},  {\em Monthly Notices of the Royal Astronomical Society} {\bf 465} (11, 2016) 4895--4913, [\href{http://arxiv.org/abs/1607.01403}{{\tt arXiv:1607.01403}}].

\bibitem{Birrer_2016}
S.~Birrer, A.~Amara, and A.~Refregier, {\it The mass-sheet degeneracy and time-delay cosmography: analysis of the strong lens {RXJ}1131-1231},  {\em Journal of Cosmology and Astroparticle Physics} {\bf 2016} (08, 2016) 020--020, [\href{http://arxiv.org/abs/1511.03662}{{\tt arXiv:1511.03662}}].

\bibitem{Birrer_2018}
S.~Birrer and A.~Amara, {\it Lenstronomy: {M}ulti-purpose gravitational lens modelling software package},  {\em Physics of the Dark Universe} {\bf 22} (Dec., 2018) 189–201, [\href{http://arxiv.org/abs/1803.09746}{{\tt arXiv:1803.09746}}].

\bibitem{Birrer_2020}
S.~Birrer, A.~J. Shajib, A.~Galan, M.~Millon, T.~Treu, A.~Agnello, M.~Auger, G.~C.-F. Chen, L.~Christensen, T.~Collett, F.~Courbin, C.~D. Fassnacht, L.~V.~E. Koopmans, P.~J. Marshall, J.-W. Park, C.~E. Rusu, D.~Sluse, C.~Spiniello, S.~H. Suyu, S.~Wagner-Carena, K.~C. Wong, M.~Barnab{\`{e} }, A.~S. Bolton, O.~Czoske, X.~Ding, J.~A. Frieman, and L.~V. de~Vyvere, {\it {TDCOSMO}: {IV}. {H}ierarchical time-delay cosmography – joint inference of the {H}ubble constant and galaxy density profiles},  {\em Astronomy \& Astrophysics} {\bf 643} (11, 2020) A165, [\href{http://arxiv.org/abs/2007.02941}{{\tt arXiv:2007.02941}}].

\bibitem{Cao_2022}
X.~Cao, R.~Li, J.~W. Nightingale, R.~Massey, A.~Robertson, C.~S. Frenk, A.~Amvrosiadis, N.~C. Amorisco, Q.~He, A.~Etherington, S.~Cole, and K.~Zhu, {\it Systematic errors induced by the elliptical power-law model in galaxy{\textendash}galaxy strong lens modeling},  {\em Research in Astronomy and Astrophysics} {\bf 22} (02, 2022) 025014, [\href{http://arxiv.org/abs/2110.14554}{{\tt arXiv:2110.14554}}].

\bibitem{Chen_2016}
C.-Y. {Chen}, C.-Y. {Hwang}, and C.-M. {Ko}, {\it {Ellipticities of Elliptical Galaxies in Different Environments}},  {\em \apj} {\bf 830} (Oct., 2016) 123.

\bibitem{teyssier2002cosmological}
R.~{Teyssier}, {\it {Cosmological hydrodynamics with adaptive mesh refinement. A new high resolution code called RAMSES}},  {\em AAP} {\bf 385} (Apr., 2002) 337--364, [\href{http://arxiv.org/abs/astro-ph/0111367}{{\tt astro-ph/0111367}}].

\bibitem{guillet2011simple}
T.~{Guillet} and R.~{Teyssier}, {\it {A simple multigrid scheme for solving the Poisson equation with arbitrary domain boundaries}},  {\em Journal of Computational Physics} {\bf 230} (June, 2011) 4756--4771, [\href{http://arxiv.org/abs/1104.1703}{{\tt arXiv:1104.1703}}].

\bibitem{komatsu2011seven}
E.~{Komatsu}, K.~M. {Smith}, J.~{Dunkley}, C.~L. {Bennett}, B.~{Gold}, G.~{Hinshaw}, N.~{Jarosik}, D.~{Larson}, M.~R. {Nolta}, L.~{Page}, D.~N. {Spergel}, M.~{Halpern}, R.~S. {Hill}, A.~{Kogut}, M.~{Limon}, S.~S. {Meyer}, N.~{Odegard}, G.~S. {Tucker}, J.~L. {Weiland}, E.~{Wollack}, and E.~L. {Wright}, {\it {Seven-year Wilkinson Microwave Anisotropy Probe (WMAP) Observations: Cosmological Interpretation}},  {\em ApJs} {\bf 192} (Feb., 2011) 18, [\href{http://arxiv.org/abs/1001.4538}{{\tt arXiv:1001.4538}}].

\bibitem{Breton:2018wzk}
M.-A. Breton, Y.~Rasera, A.~Taruya, O.~Lacombe, and S.~Saga, {\it {Imprints of relativistic effects on the asymmetry of the halo cross-correlation function: from linear to non-linear scales}},  {\em Mon. Not. Roy. Astron. Soc.} {\bf 483} (2019), no.~2 2671--2696, [\href{http://arxiv.org/abs/1803.04294}{{\tt arXiv:1803.04294}}].

\bibitem{reverdy2014propagation}
V.~Reverdy, {\em Propagation de la lumière dans un Univers structuré et nouvelles approches numériques en cosmologie}.
\newblock PhD thesis, Laboratoire Univers et Th\'eories, 2014.
\newblock \href{http://arxiv.org/abs/https://theses.hal.science/tel-02095297v1}{{\tt https://theses.hal.science/tel-02095297v1}}.
\newblock \url{https://github.com/vreverdy/magrathea-pathfinder/blob/master/vreverdy_phd_manuscript.pdf}.

\bibitem{Breton}
M.-A. Breton, {\em Constructing observables in cosmology : towards new probes of the dark sector}.
\newblock PhD thesis, Laboratoire Univers et Th\'eories, 2018.
\newblock \href{http://arxiv.org/abs/https://theses.hal.science/tel-02114661/}{{\tt https://theses.hal.science/tel-02114661/}}.

\bibitem{Bovy_2023}
J.~Bovy, {\em Dynamics and Astrophysics of Galaxies}.
\newblock Princeton University Press (in preparation), 2023.

\bibitem{Pizzella_2005}
A.~Pizzella, E.~M. Corsini, E.~Dalla~Bonta, M.~Sarzi, L.~Coccato, and F.~Bertola, {\it On the relation between circular velocity and central velocity dispersion in high and low surface brightness galaxies},  {\em The Astrophysical Journal} {\bf 631} (Oct., 2005) 785–791, [\href{http://arxiv.org/abs/astro-ph/0503649}{{\tt astro-ph/0503649}}].

\bibitem{Corsini_2007}
E.~Corsini, A.~Pizzella, E.~D. Bont{\`a}, F.~Bertola, L.~Cocato, and M.~Sarzi, {\it Circular velocity and central velocity dispersion in low surface brightness galaxies},  in {\em ISLAND UNIVERSES} (R.~S. DE~JONG, ed.), (Dordrecht), pp.~77--82, Springer Netherlands, 2007.

\bibitem{Kochanek_2002}
C.~S. Kochanek, {\it What do gravitational lens time delays measure?},  {\em The Astrophysical Journal} {\bf 578} (10, 2002) 25--32.

\bibitem{Sonnenfeld_2017}
A.~Sonnenfeld, {\it On the choice of lens density profile in time delay cosmography},  {\em Monthly Notices of the Royal Astronomical Society} {\bf 474} (12, 2017) 4648--4659.

\bibitem{Birrer_2021}
S.~Birrer, {\it Gravitational lensing formalism in a curved arc basis: A continuous description of observables and degeneracies from the weak to the strong lensing regime},  {\em The Astrophysical Journal} {\bf 919} (Sept., 2021) 38, [\href{http://arxiv.org/abs/2110.14554}{{\tt arXiv:2110.14554}}].

\bibitem{Bartelmann_1998}
M.~Bartelmann, A.~Huss, J.~M. Colberg, A.~Jenkins, and F.~R. Pearce, {\it Arc statistics with realistic cluster potentials. {IV}. clusters in different cosmologies},  {\em Astronomy \& Astrophysics} (1997) [\href{http://arxiv.org/abs/astro-ph/9707167}{{\tt astro-ph/9707167}}].

\bibitem{Meneghetti_2013}
M.~Meneghetti, M.~Bartelmann, H.~Dahle, and M.~Limousin, {\it Arc statistics},  {\em Space Science Reviews} {\bf 177} (05, 2013) 31--74, [\href{http://arxiv.org/abs/1303.3363}{{\tt arXiv:1303.3363}}].

\bibitem{Puchwein_2009}
E.~Puchwein and S.~Hilbert, {\it Cluster strong lensing in the millennium simulation: the effect of galaxies and structures along the line-of-sight},  {\em Monthly Notices of the Royal Astronomical Society} {\bf 398} (2009), no.~3 1298--1308, [\href{http://arxiv.org/abs/0904.0253}{{\tt arXiv:0904.0253}}].

\bibitem{Bayliss_2014}
M.~B. Bayliss, T.~Johnson, M.~D. Gladders, K.~Sharon, and M.~Oguri, {\it Line-of-sight structure toward strong lensing galaxy clusters},  {\em The Astrophysical Journal} {\bf 783} (02, 2014) 41, [\href{http://arxiv.org/abs/1312.3637}{{\tt arXiv:1312.3637}}].

\bibitem{wells-2024}
P.~R. Wells, C.~D. Fassnacht, S.~Birrer, and D.~Williams, {\it {TDCOSMO} {XVI}: {P}opulation analysis of lines of sight of 25 strong galaxy-galaxy lenses with extreme value statistics},  2024.

\end{thebibliography}\endgroup
